\documentclass[11pt,a4paper]{article}
\usepackage{jheppub}
\usepackage{url}
\usepackage{graphicx}

\begin{document}


\title{Search for magnetic monopoles with the MoEDAL prototype trapping detector in 8 TeV proton-proton collisions at the LHC}

\author[]{THE MoEDAL COLLABORATION: \\}
\author[a,b]{B.~Acharya,}
\author[a]{J.~Alexandre,}
\author[v]{K.~Bendtz,}
\author[k]{P.~Benes,}
\author[c]{J.~Bernab\'{e}u,}
\author[da]{M.~Campbell,}
\author[w]{S.~Cecchini,}
\author[f]{J.~Chwastowski,}
\author[l]{A.~Chatterjee,}
\author[g]{M.~de~Montigny,}
\author[f]{D.~Derendarz,}
\author[da]{A.~De~Roeck,}
\author[a,db]{J.~R.~Ellis,}
\author[a]{M.~Fairbairn,}
\author[h]{D.~Felea,}
\author[ia]{M.~Frank,}
\author[j]{D.~Frekers,}
\author[c]{C.~Garcia,}
\author[e,\dagger]{G.~Giacomelli\note[\textdagger]{Now deceased},}
\author[ia]{D.~Ha\c{s}egan,}
\author[dc]{M.~Kalliokoski,}
\author[l]{A.~Katre,}
\author[m]{D.-W.~Kim,}
\author[c]{M.~G.~L.~King,}
\author[n]{K.~Kinoshita,}
\author[da]{D.H.~Lacarr\`ere,}
\author[m]{S.~C.~Lee,}
\author[ib]{C.~Leroy,}
\author[l]{A.~Lionti,}
\author[e]{A.~Margiotta,}
\author[w]{N.~Mauri,}
\author[a]{N.~E.~Mavromatos,}
\author[l,1]{P.~Mermod\note{Communicating author. Email: philippe.mermod@cern.ch},}
\author[v]{D.~Milstead,}
\author[c]{V.~A.~Mitsou,}
\author[o]{R.~Orava,}
\author[y]{B.~Parker,} 
\author[e]{L.~Pasqualini,}
\author[w]{L.~Patrizii,}
\author[h]{G.~E.~P\u{a}v\u{a}la\c{s},}
\author[g]{J.~L.~Pinfold,}
\author[k]{M.~Platkevi\v{c},}
\author[h]{V.~Popa,}
\author[w]{M.~Pozzato,}
\author[k]{S.~Pospisil,}
\author[p]{A.~Rajantie,}
\author[w,q]{Z.~Sahnoun,}
\author[a]{M.~Sakellariadou,}
\author[a]{S.~Sarkar,}
\author[r]{G.~Semenoff,}
\author[w]{G.~Sirri,}
\author[s]{K.~Sliwa,}
\author[g]{R.~Soluk,}
\author[e]{M.~Spurio,}
\author[t]{Y.~N.~Srivastava,}
\author[f]{R.~Staszewski,}
\author[k]{M.~Suk,}
\author[t]{J.~Swain,}
\author[x]{M.~Tenti,}
\author[w]{V.~Togo,}
\author[f]{M.~Trzebinski,}
\author[g]{J.~A.~Tuszy\'{n}ski,}
\author[c]{V.~Vento,}
\author[c]{O.~Vives,}
\author[k]{Z.~Vykydal,}
\author[y,z]{T.~Whyntie,} 
\author[t]{A.~Widom,}
\author[j]{G.~Willems,}
\author[u]{and J.~H.~Yoon} 

\affiliation[a]{Theoretical Particle Physics \& Cosmology Group, Physics Dept., King's College London, UK}
\affiliation[b]{International Centre for Theoretical Physics, Trieste, Italy}
\affiliation[c]{IFIC, Universitat de Val\`{e}ncia - CSIC, Valencia, Spain}
\affiliation[da]{Experimental Physics Department, CERN, Geneva, Switzerland}
\affiliation[db]{Theoretical Physics Department, CERN, Geneva, Switzerland}
\affiliation[dc]{Beams Department, CERN, Geneva, Switzerland}
\affiliation[e]{INFN, Section of Bologna \& Department of Physics \& Astronomy, University of Bologna, Italy}
\affiliation[f]{Institute of Nuclear Physics Polish Academy of Sciences, Cracow, Poland}
\affiliation[g]{Physics Department, University of Alberta, Edmonton, Alberta, Canada}
\affiliation[h]{Institute of Space Science, Bucharest - M\u{a}gurele, Romania}
\affiliation[ia]{Department of Physics, Concordia University, Montr\'{e}al, Qu\'{e}bec,  Canada}
\affiliation[ib]{D\'{e}partement de physique, Universit\'{e} de Montr\'{e}al, Qu\'{e}bec,  Canada}
\affiliation[j]{Physics Department, University of Muenster, Muenster, Germany}
\affiliation[k]{IEAP, Czech Technical University in Prague, Czech~Republic}
\affiliation[l]{Section de Physique, Universit\'{e} de Gen\`{e}ve, Geneva, Switzerland}
\affiliation[m]{Physics Department, Gangneung-Wonju National University, Gangneung, Republic of Korea}
\affiliation[n]{Physics Department, University of Cincinnati, Cincinnati, Ohio, USA}
\affiliation[o]{Physics Department, University of Helsinki, Helsinki, Finland}
\affiliation[p]{Department of Physics, Imperial College London, UK}
\affiliation[q]{Centre for Astronomy, Astrophysics and Geophysics, Algiers, Algeria}
\affiliation[r]{Department of Physics, University of British Columbia, Vancouver, British Columbia, Canada}
\affiliation[s]{Department of Physics and Astronomy, Tufts University, Medford, Massachusetts, USA}
\affiliation[t]{Physics Department, Northeastern University, Boston, Massachusetts, USA}
\affiliation[u]{Physics Department, Konkuk University, Seoul, Korea}
\affiliation[v]{Physics Department, Stockholm University, Stockholm, Sweden - Associate member}
\affiliation[w]{INFN, Section of Bologna, Bologna, Italy}
\affiliation[x]{INFN, CNAF, Bologna, Italy}
\affiliation[y]{The Institute for Research in Schools, Canterbury, England}
\affiliation[z]{Queen Mary University of London, London, England}

\dedicated{This paper is dedicated to the memory of Giorgio Giacomelli, a pioneer and leader in the quest for the Magnetic Monopole.}

\date{\today}

\abstract{The MoEDAL experiment is designed to search for magnetic monopoles and other highly-ionising particles produced in high-energy collisions at the LHC. The largely passive MoEDAL detector, deployed at Interaction Point 8 on the LHC ring, relies on two dedicated direct detection techniques. The first technique is based on stacks of nuclear-track detectors with surface area $\sim$18 m$^2$, sensitive to particle ionisation exceeding a high threshold. These detectors are analysed offline by optical scanning microscopes. The second technique is based on the trapping of charged particles in an array of roughly 800 kg of aluminium samples. These samples are monitored offline for the presence of trapped magnetic charge at a remote superconducting magnetometer facility. We present here the results of a search for magnetic monopoles using a 160 kg prototype MoEDAL trapping detector exposed to 8 TeV proton-proton collisions at the LHC, for an integrated luminosity of 0.75 fb$^{-1}$. No magnetic charge exceeding $0.5g_{\rm D}$ (where $g_{\rm D}$ is the Dirac magnetic charge) is measured in any of the exposed samples, allowing limits to be placed on monopole production in the mass range 100~GeV$\leq m \leq$ 3500~GeV. Model-independent cross-section limits are presented in fiducial regions of monopole energy and direction for $1g_{\rm D}\leq|g|\leq 6g_{\rm D}$, and model-dependent cross-section limits are obtained for Drell-Yan pair production of spin-1/2 and spin-0 monopoles for $1g_{\rm D}\leq|g|\leq 4g_{\rm D}$. Under the assumption of Drell-Yan cross sections, mass limits are derived for $|g|=2g_{\rm D}$ and $|g|=3g_{\rm D}$ for the first time at the LHC, surpassing the results from previous collider experiments. }

\keywords{new physics, high-energy collisions, LHC, magnetic monopole, SQUID magnetometer, persistent current}

\maketitle

\section{Introduction}
\label{intro}

Once the unification of electricity and magnetism was achieved by James Clerk Maxwell~\cite{Maxwell1865}, it was  natural to hypothesise the existence of isolated magnetic charge --- the counterparts of  electric charge --- that would symmetrise Maxwell's equations and complete the electric-magnetic duality. One of the first to recognise that  magnetic monopoles could conceivably exist was Pierre Curie~\cite{Curie1894}. Interest in the existence of magnetic monopoles greatly increased when Dirac showed that electric charge quantisation could be explained as a natural consequence of angular momentum quantisation in the presence of a monopole~\cite{Dirac1931}. A paradigm shift in our understanding of magnetic monopoles occurred in 1974 when 't Hooft and Polyakov independently showed that certain spontaneously-broken gauge theories --- where the U(1) subgroup of electromagnetism is embedded into a gauge group that is spontaneously broken by the Higgs mechanism --- necessarily possess a topological magnetic monopole solution~\cite{tHooft1974,Polyakov1974}. Thus monopoles are now not just a possibility but a prediction of such theories. More recently it has been proposed that magnetic monopole solutions could arise within the electroweak theory itself~\cite{Cho1997}. This Cho-Maison or electroweak monopole would be expected to have a mass of the order of several TeV~\cite{Kirkman1981,Cho2015,Ellis2016}, possibly within reach of the LHC. 
 
It follows from Dirac's argument that the magnetic charge $q_m$ carried by a monopole should be a multiple of the fundamental Dirac magnetic charge. In Gaussian units, the Dirac quantisation relation reads:    

\begin{equation}
g=\frac{q_m}{e}=\frac{n}{2\alpha_{\rm e}} = n\cdot g_{\rm D} \approx n\cdot 68.5,
\label{eqn:gcharge}
\end{equation}
where $e$ is the electric charge of the proton, $\alpha_{\rm e}$ is the fine structure constant, $n$ is an integer and $g_D$ is the Dirac unit of magnetic charge. In SI units, the dimensionless quantity $g$~\footnote{There exist alternative definitions for the magnetic charge number $g$ in the literature (see~\cite{MoEDAL2014} and references therein). The definition chosen here allows $g$ to enter the Bethe formula in the same way as the charge number $z$ for electric charge (see Equation~\ref{Bethe_mag}).} is related to the magnetic charge $q_m$ by the relation $q_m=gec$ where $c$ is the speed of light in vacuo. 

The large value of the Dirac charge, $g_{\rm D}=68.5$, implies that the minimum coupling of a monopole to the photon should be much larger than 1, preventing reliable perturbative calculations of monopole production processes. According to Dirac's theory, the minimum value of the magnetic charge is $g_{\rm D}$ if the fundamental charge is $e$, and would become $3g_{\rm D}$ if free particles with charge $\frac{1}{3}e$ were to exist~\cite{Preskill1984}. Schwinger proposed a magnetic model of matter based on dyons whose magnetic charge is at least $2g_{\rm D}$~\cite{Schwinger1975}. The minimum magnetic charge is $g_{\rm D}$ or $2g_{\rm D}$ when the monopole is a topological object such as the grand-unification monopole~\cite{tHooft1974} and it is $2g_{\rm D}$ for the electroweak monopole~\cite{Cho1997}. In all cases the lightest magnetic monopole is stable by virtue of magnetic charge conservation. The in-flight ionisation energy loss of a relativistic monopole with a single Dirac charge is equivalent to that of a relativistic electrically-charged particle with $|z|\approx 68.5$, where $z$ is the number of elementary charges $e$. Thus, a relativistic monopole would induce an ionisation in matter over $4700$ times higher than a minimum-ionising particle~\cite{Ahlen1978,Ahlen1980,Ahlen1982}. 

Accelerator-based experiments designed to search for magnetic monopoles are carried out at each advance of the high-energy frontier~\cite{Fairbairn2007,Patrizii2015}. Astroparticle experiments have also played a key role in the monopole quest at the cosmic frontier~\cite{Burdin2015}. At colliders, monopoles would be produced in pairs and manifest themselves as very highly-ionising particles. The monopoles produced would quickly slow down and get trapped in the material surrounding the interaction points. An indisputable test for the presence of a trapped magnetic monopole in matter would be the measurement of a persistent current induced when a sample is passed through the superconducting coil of a SQUID magnetometer~\cite{Fairbairn2007,Burdin2015}. This induction technique relies on the assumption that monopole binding to matter is stronger than the forces which are subsequently applied to the samples, such as the gravitational and magnetic pulls from the Earth. The use of ferromagnetic material, which would guarantee binding via the mirror image force~\cite{Kittel1977,Goto1963}, is not practical in the present search for two reasons: such material can cause physical damage by moving freely when exposed to magnetic fields in the LHC caverns, and its large magnetic moments can cause SQUID magnetometer instabilities. The search therefore relies on strong monopole binding to matter through interactions with atoms or nuclei. This is a reasonable assumption for nuclei with non-zero magnetic dipole moments such as aluminium, which are expected to bind with monopoles with binding energies of the order of hundreds of keV~\cite{Milton2006}.  


Three kinds of techniques were used to detect magnetic monopoles in previous collider experiments  \cite{Fairbairn2007}. The first relies on the detection of energy loss, either using active detectors such as gaseous ionisation chambers and scintillators  --- as employed, for example, by OPAL at LEP~\cite{OPAL2008} and CDF at the Tevatron~\cite{CDF2006} --- or passive plastic Nuclear Track Detectors (NTDs) --- as employed, for example, at the Tevatron~\cite{Bertani1990} and by MODAL~\cite{Kinoshita1992} and OPAL~\cite{Pinfold1993} at LEP. The second method utilises the recognition of the parabolic trajectory signature of a monopole in a magnetic field --- for example, in TASSO~\cite{TASSO1988} and OPAL~\cite{Pinfold1991}. The third method applies the induction technique to material adjacent to the interaction point in which some monopoles can range out and be trapped. This search method was employed, for example, using exposed detector and accelerator material at HERA~\cite{H12005} and at the Tevatron~\cite{Kalbfleisch2000,Kalbfleisch2004}. Experiments using all three methods limited the production of particles with magnetic charges equal to or above the Dirac charge and masses up to $400$ GeV.

Masses larger by one order of magnitude can be probed at the LHC in high-energy proton-proton ($pp$) collisions, using the same kinds of techniques~\cite{DeRoeck2012a}. Searches for highly-ionising particles performed at the ATLAS general-purpose experiment in 7 and 8 TeV $pp$ collisions resulted in the exclusion of singly-charged magnetic monopoles with masses up to the order of 1 TeV assuming Drell-Yan cross sections extrapolated to high electromagnetic charges~\cite{ATLAS2012a,ATLAS2015a}. These results are valid for magnetic charges up to $|g|=1.5g_{\rm D}$ and electric charges up to $|z|=60$. However, certain caveats apply here. The general-purpose LHC detectors ATLAS and CMS are designed and optimised to detect relativistic electrically-charged particles and energetic neutral particles such as photons and neutrons. In the case of very highly-ionising particles the performance of these detectors has not been experimentally assessed and calibrated. Consequently, their response to highly-ionising particles is estimated using simulations.

The MoEDAL experiment~\cite{MoEDAL2009} at the LHC is dedicated to the search for highly-ionising messengers of physics beyond the Standard Model such as magnetic monopoles and massive long-lived or stable charged particles~\cite{MoEDAL2014}. MoEDAL utilises two complementary types of passive detectors: NTD stacks that are sensitive only to highly-ionising particles, and a magnetic monopole trapping detector consisting of roughly 800 kg of aluminium samples that is capable of trapping highly-ionising particles for further study. MoEDAL's radiation environment is monitored by a real-time TimePix pixel detector array. After exposure the plastic NTDs are etched and analysed using computer controlled optical scanning microscopes to search for tracks comprised of collinear etch pits pointing back to the interaction point. Exposed trapping detectors are monitored for the presence of trapped magnetic charge at a SQUID magnetometer facility. 

The use of passive detection techniques means that MoEDAL is free of the assumptions and restrictions of an electronic trigger as well as the limitations of electronics readout. MoEDAL's relatively open intersection region at Interaction Point 8 (IP8) results in a low material budget for the passage of highly-ionising particles. Importantly, the MoEDAL detector response to highly-ionising particles is directly calibrated using heavy-ion beams. Another crucial feature of the MoEDAL detector is the absence of backgrounds. Standard Model particles may induce reactions within the NTD plates that fog the image but can never simulate a signal of new physics. Likewise, there is no possible way that Standard Model backgrounds can mimic the signal of a trapped monopole. MoEDAL can directly measure magnetic charge in a robust way. The complementary nature of the MoEDAL detector, its minimisation of experimental assumptions and its well understood response --- exemplified by its ability to retain a permanent record, and even capture new particles for further study --- makes MoEDAL an invaluable asset in the elucidation of any Terascale new physics scenario covered by its extensive physics programme~\cite{MoEDAL2014}. 

In this paper, we present results from the MoEDAL trapping detector prototype deployed in 2012 and exposed to 8 TeV $pp$ collisions. The paper is structured as follows. The MoEDAL prototype trapping detector setup is described in Section~\ref{experiment}. The experimental method utilised for the SQUID magnetometer measurements as well as  results of the magnetometer scans are presented in Section~\ref{magnetometer}. Section~\ref{simulations} describes how monopole propagation in matter is simulated. The trapping acceptance of the detector for various values of monopole mass and charge is extracted and discussed in Section~\ref{monopole_trapping_acceptance}. The limits obtained on monopole production are then presented in Section~\ref{limits}. Finally, the conclusions and outlook are presented in Section~\ref{conclusions}. 

\section{The MoEDAL prototype trapping detector}
\label{experiment}

The MoEDAL detector described above is deployed around IP8 on the LHC ring, in the VErtex LOcator (VELO)~\cite{LHCb2008} cavern of the LHCb experiment. The MoEDAL trapping detector prototype consists of 160 kg of aluminium rods of 60 cm length and 2.5 cm diameter housed in 11 boxes, with each box comprising 18 rods. It was deployed immediately underneath the beam pipe directly upstream of the LHCb VELO detector vacuum vessel. Aluminium is a particularly good choice for the trapping volume material for four important reasons. First, the large magnetic moment of the $^{27}_{13}$Al nucleus means that it is expected to bind strongly with a monopole, thus capturing it inside the atomic lattice~\cite{Milton2006} (see further discussion in Section~\ref{monopole_trapping_criterion}). Secondly, aluminium does not present a problem with respect to activation. Thirdly, aluminium is non-magnetic, which favours the stability of the SQUID magnetometer during measurements. Last but not least, it presents a readily available and cost-effective solution. 

\begin{figure}[tb]
\begin{center}
  \includegraphics[width=0.495\linewidth]{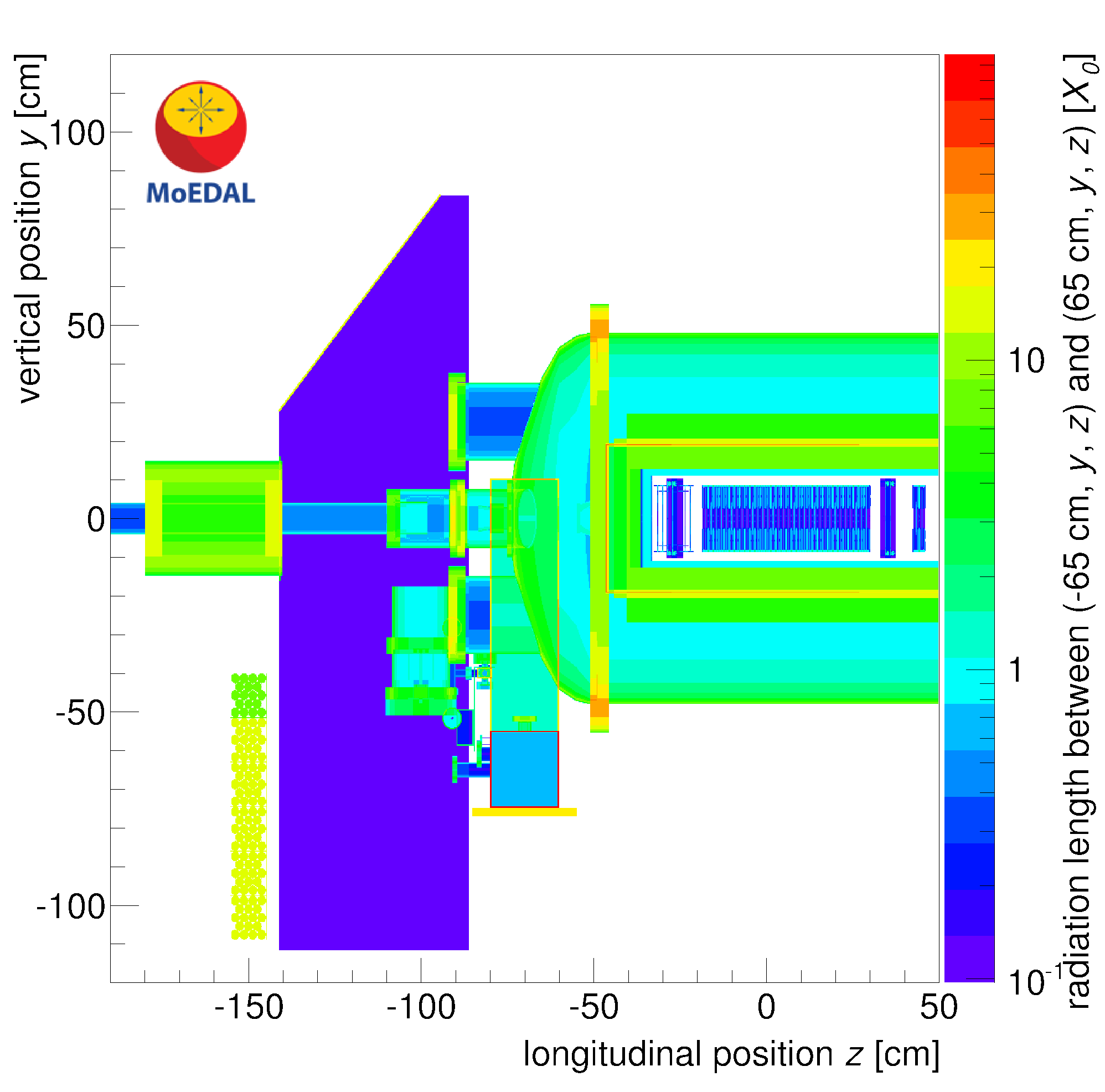}
  \includegraphics[width=0.495\linewidth]{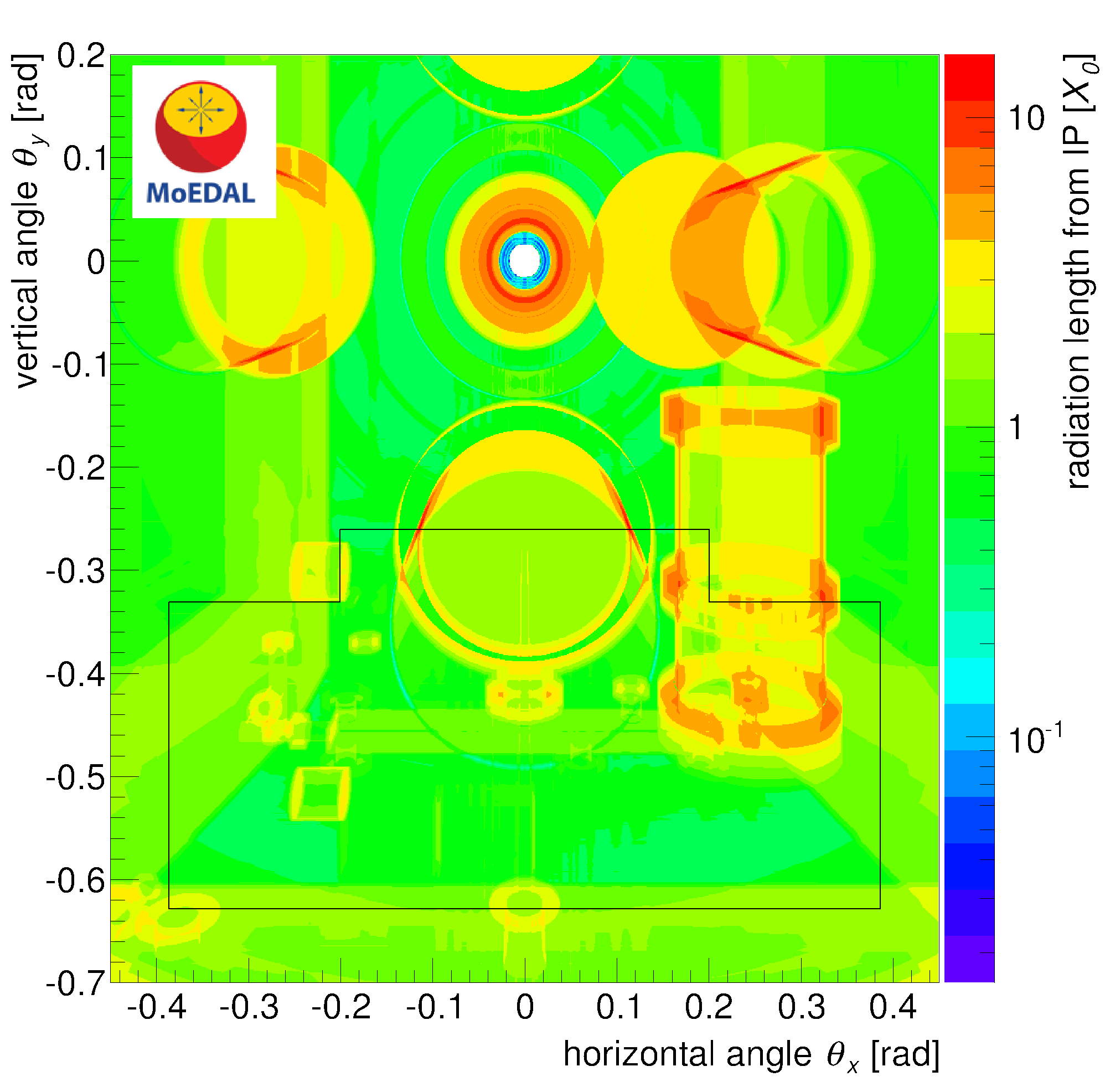}  
  \caption{Material budget in radiation length in the $yz$ plane for $|x|<65$ cm (left) and in the $\theta_x\theta_y$ plane for $z=-145$ cm (right). In the right figure, the outline corresponding to the trapping detector position is indicated in black. The grid used as an approximation to model cables and pipes is not included in these figures.}
\label{fig:MMTgeo}
\end{center}
\end{figure}

The boxes comprising the trapping detector were stacked in two columns and numbered from 1 to 11 starting from the bottom, with the eleventh box centred on top of the two columns. The position of the centre of the top box in the LHCb coordinate system\footnote{A right-handed coordinate system in which the $z$ axis points along the beam in the direction of the LHCb detectors, $y$ is the vertical direction, and $x$ is the horizontal direction is used.} was ($x$,$y$,$z$)=(0,$-$45 cm,$-$150 cm) with an uncertainty of 1 cm for each coordinate. 
Fig.~\ref{fig:MMTgeo} summarises the geometry of the detector and its surroundings and quantifies the amount of material in radiation lengths ($X_0$) present in the installation. The material budget between the interaction point and the trapping detector varied from 0.1 to 8 $X_0$ depending on position, with an average of 1.4 $X_0$. 

The main contributions to the material arise from the VELO vacuum vessel interior and outer wall, various flanges, a vacuum pump, and vacuum manifold components attached to the VELO vessel. These elements were implemented in a geometry model, using the LHCb geometry model as a basis. In order to model cables and small pipes in the region, the approximation of a grid of material was used, with 101 vertical steel rods of radius 0.3 cm, spaced out by 1 cm at $z=-115$ cm. This represents on average 2.3\% of the total radiation length. To model material uncertainties, two additional material geometries were utilised, based on conservative estimates of the maximum and minimum amount of material that could be plausibly unaccounted for due to uncertainties in the material budget. These geometries were implemented by changing the grid rod radius to 0.01 cm (minimum extra material) and 0.5 cm (maximum extra material), respectively. This dominating systematic uncertainty on the trapping detector acceptance is estimated by propagating monopoles in matter in the different geometries using the G\textsc{eant}4 toolkit~\cite{Geant42006} (see Section~\ref{simulations}).

The MoEDAL prototype trapping detector was exposed to 8 TeV $pp$ collisions at IP8 between September and December 2012. Using LHCb luminosity measurements during that period, conservatively assuming a 3.5\% uncertainty~\cite{LHCb2014}, this corresponds to an integrated luminosity of $0.75\pm 0.03$ fb$^{-1}$. After the run was finished, the rods were retrieved and cut into samples of 20 cm length with a non-ferromagnetic saw (except for the top box, whose rods were cut into a mix of 10, 15, 20 and 30 cm samples for studying the sample-size dependence of the magnetometer response), for a total of 606 samples.

\section{Magnetometer measurements}
\label{magnetometer}

A DC-SQUID rock magnetometer (2G Enterprises model 755) housed at the Laboratory for Natural Magnetism at ETH Zurich was used to scan the trapping detector samples. The magnetometer calibration was performed with a convolution method applied to a dipole sample, and cross-checked using long thin solenoids that mimic a monopole of well-known magnetic charge~\cite{DeRoeck2012b}. Two 25 cm long solenoids of different coil areas and different number of turns were used with currents varying from 0.01 to 10 $\mu$A. The magnetometer response was measured to be linear and charge symmetric in a range corresponding to $0.1-300g_{\rm D}$. After calibration, the measured current is translated into units of current expected from the passage of a Dirac magnetic charge, $I_{g_{\rm D}}$, with an estimated uncertainty of 10\% in the calibration constant as obtained by comparing the different independent methods~\cite{DeRoeck2012b}. A calibration dipole sample of well-known magnetic moment was measured at the start of each run, with current values found to be consistent within 1\%, ensuring that the calibration constant remained stable. Each of the 606 aluminium samples of the trapping detector was passed at least once through the magnetometer, mostly during a measurement campaign in September 2013. Roughly every tenth measurement was performed with an empty sample holder for offset subtraction. 

\begin{figure}[tb]
\begin{center}
  \includegraphics[width=0.49\linewidth]{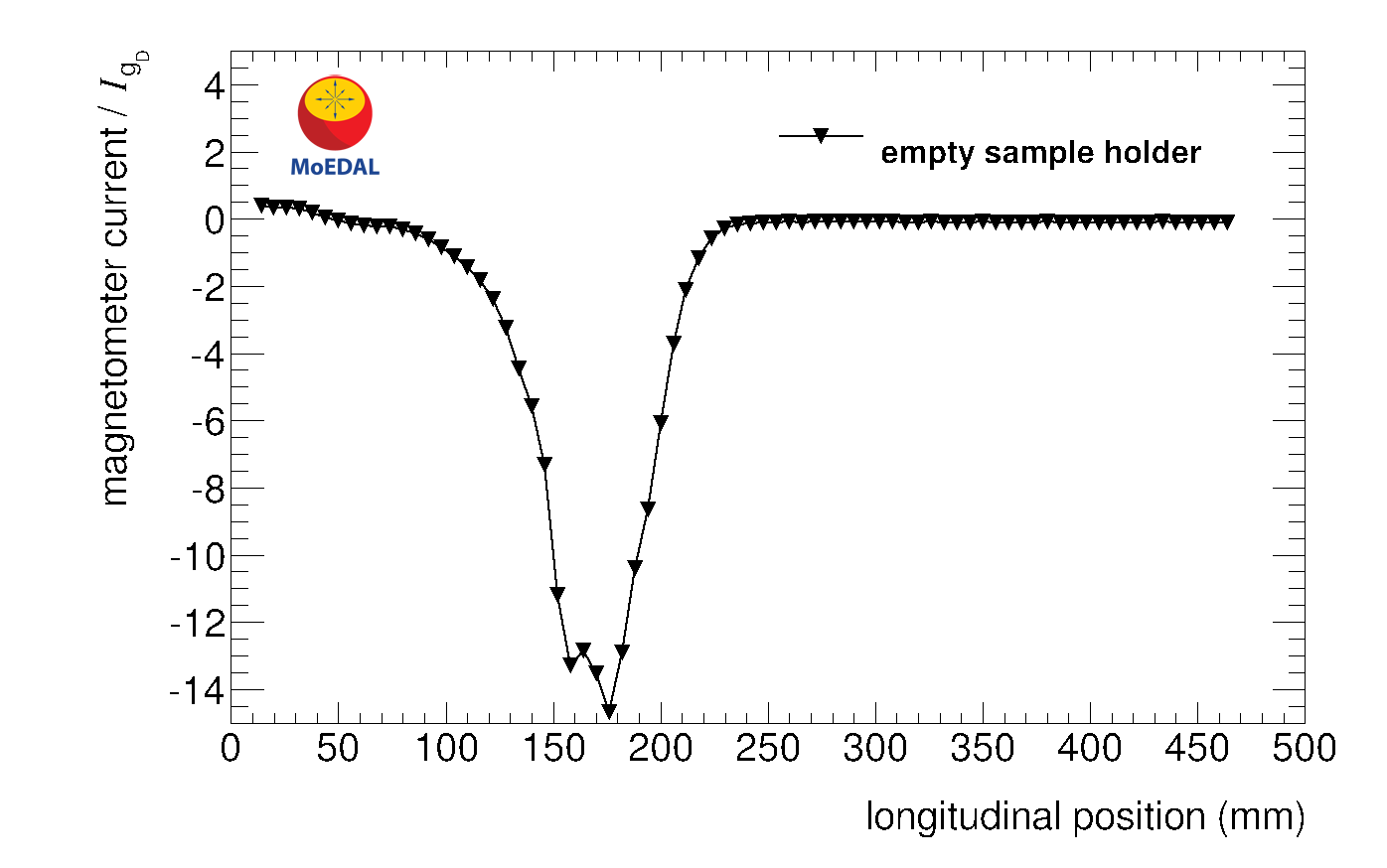}
  \includegraphics[width=0.49\linewidth]{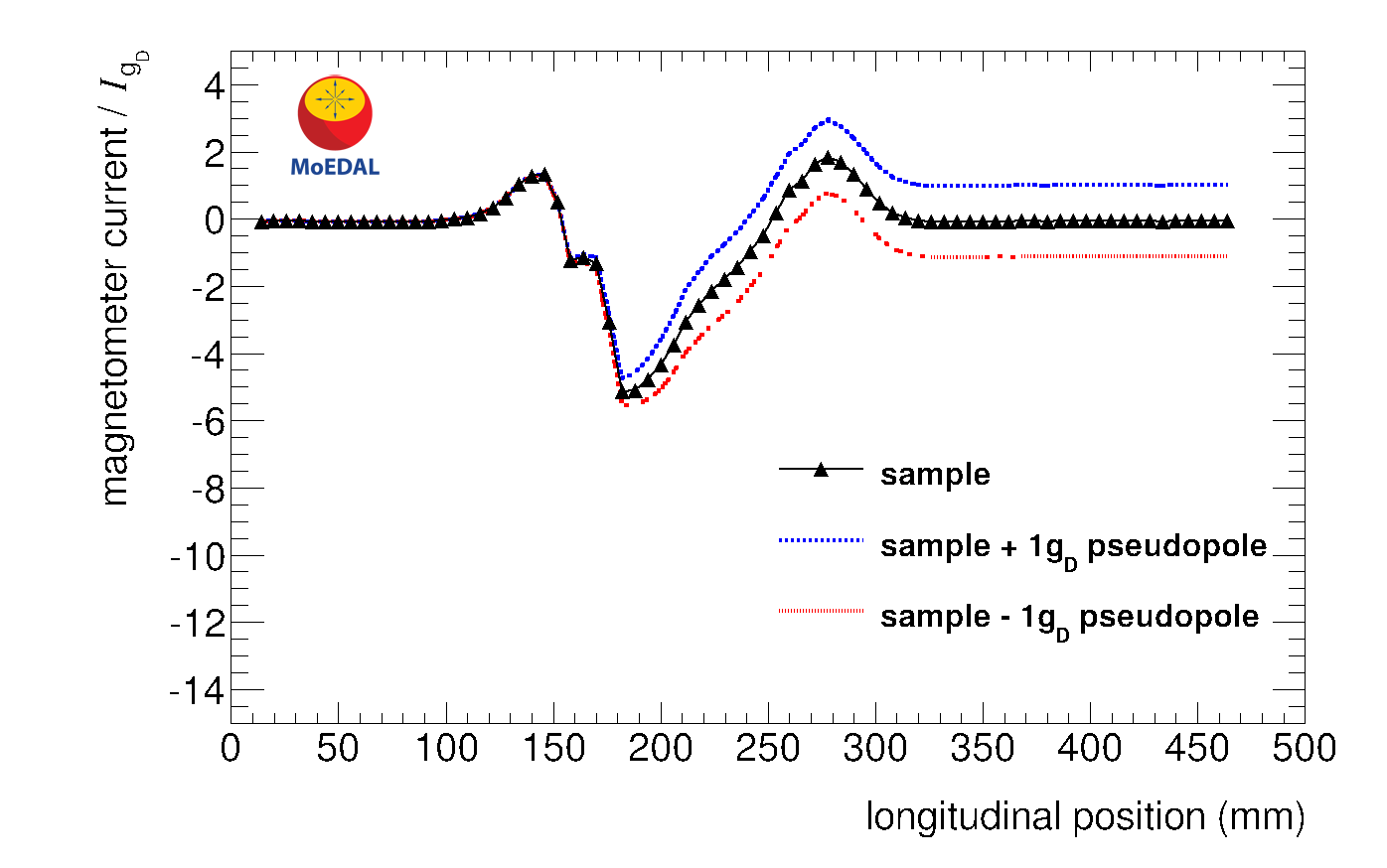}
  \caption{Magnetometer response profile along the axis of the tube for the empty sample holder (left) and a typical aluminium sample of the trapping detector after subtracting the response of the empty sample holder (right). The 20 cm sample is inside the sensing region for longitudinal positions between 110 mm and 310 mm. Solid lines simply connect the points. Dashed lines show the responses when the measurement from a long solenoid is added and subtracted to emulate the presence of a Dirac monopole in the sample.}
\label{fig:MMTprofile}
\end{center}
\end{figure}

The SQUID magnetometer's sample holder is a long telescopic carbon-fibre tube which extends completely along the axis of the sensing coils when it is at the start position. The sample is placed at the position of maximum extension and then slowly moved back through the sensing coils to a final position of minimum extension of the sample holder arm. Magnetic dipole impurities inside the sample and the sample holder can induce currents in the coils. Fig.~\ref{fig:MMTprofile} shows measurements with one 20 cm sample at 76 different positions before, during and after passage through the sensing coils, after subtracting the same measurements with an empty sample holder. It can be noted that the magnetisation of the sample holder itself is of the same order as the typical sample magnetisation. This provides an example of a typical magnetometer response profile as a function of sample position. An emulation of the response expected if a positive or negative monopole was present in the sample is given in the figure by adding or subtracting measurements obtained with a long solenoid scaled to the current expected from a Dirac monopole $I_{g_{\rm D}}$. With a monopole present a persistent current substantially different from zero would be recorded. The monopole signature is therefore measured in terms of the persistent current, defined as the difference between the currents measured after and before passage of the sample through the sensing coil, after adjustment for the contribution of the sample holder.

\begin{figure}[tb]
\begin{center}
  \includegraphics[width=0.99\linewidth]{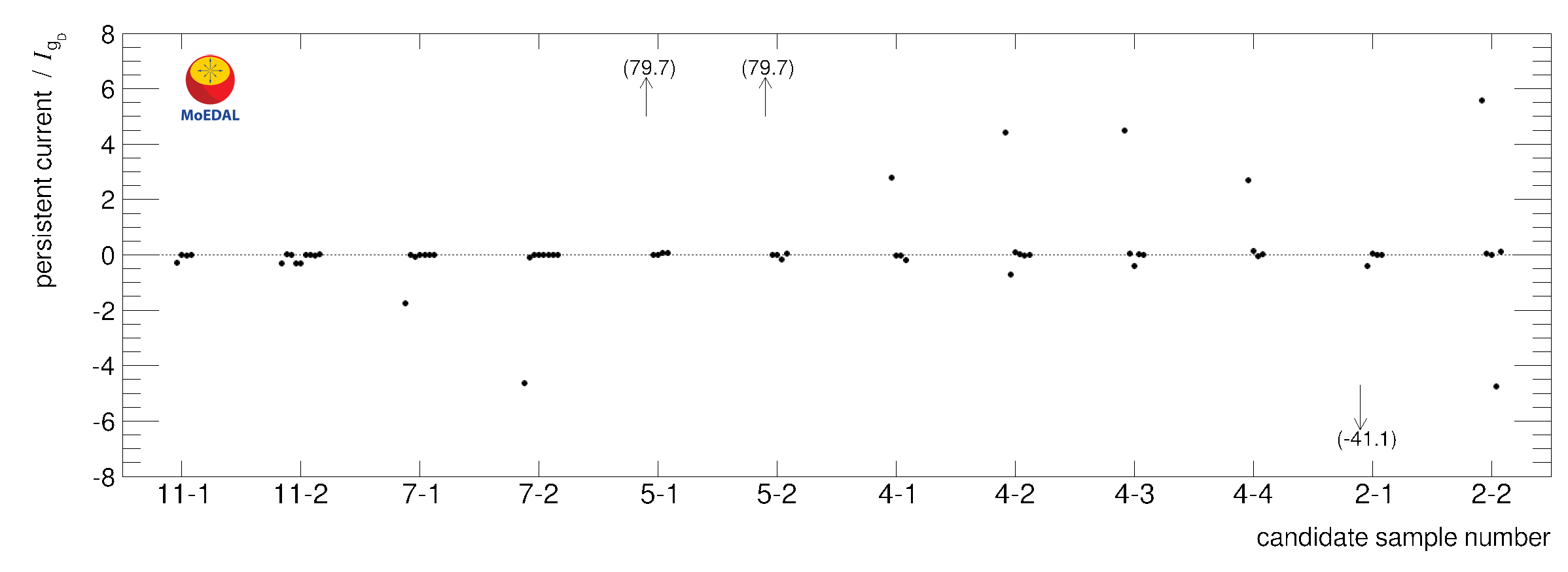}
  \caption{Results of multiple persistent current measurements (in units of the Dirac charge) for the 12 samples which yielded large ($|g|>0.25~g_{\rm D}$) values for the first measurement. Repeated measured values consistent with zero magnetic charge show that the first measurement was affected by a spurious jump. The arrows indicate values which lie off the scale of the plot.}
\label{fig:MMcandidates}
\end{center}
\end{figure}

Whenever the persistent current differed from zero by more than $0.25g_{\rm D}$ the sample was considered a candidate and measured again several times. Spurious jumps caused this to happen in $\sim$2\% of the measurements.  A sample containing a genuine monopole would consistently yield the same value for repeated measurements, while values repeatedly consistent with zero would be measured when no monopole is present. Multiple measurements of potential candidates are shown in Fig.~\ref{fig:MMcandidates}. In all cases where the first measurement showed a large fluctuation, subsequent additional check measurements of the same sample were consistent with zero. 

It was noticed that jumps occurred more often for certain periods during which the magnetometer response was less stable than usual. Instabilities can be caused by several known instrumental and environmental factors, including: spurious flux jumps occurring when the slew rate is increased~\cite{Clarke2006} as, for instance, when a sample contains ferromagnetic impurities or when a large sample is passed through the sensing coil at a high speed; noise currents in the SQUID feedback loop; the accumulation of condensed water and ice in the magnetometer tube near the cold sensing region; physical vibrations and shocks; small ($\sim$mm) variations in the initial position of the sample holder from one run to another; and variations in external magnetic fields, in particular the geomagnetic field but also possibly fields from high-voltage power line activity in the vicinity of the laboratory. Smooth variations in external magnetic fields also cause the current offset to drift, typically by a value corresponding to $0.05g_{\rm D}$ in the course of one hour. Precautions taken at regular intervals in order to minimise SQUID coil current jumps and drifts include:  cleaning the SQUID magnetometer tube through which the sample and sample holder pass; resetting the SQUID current offset; and performing frequent empty holder measurements. Such safeguards help mitigate the impact of systematic effects on the magnetic charge resolution. 

\begin{figure}[tb]
\begin{center}
  \includegraphics[width=0.99\linewidth]{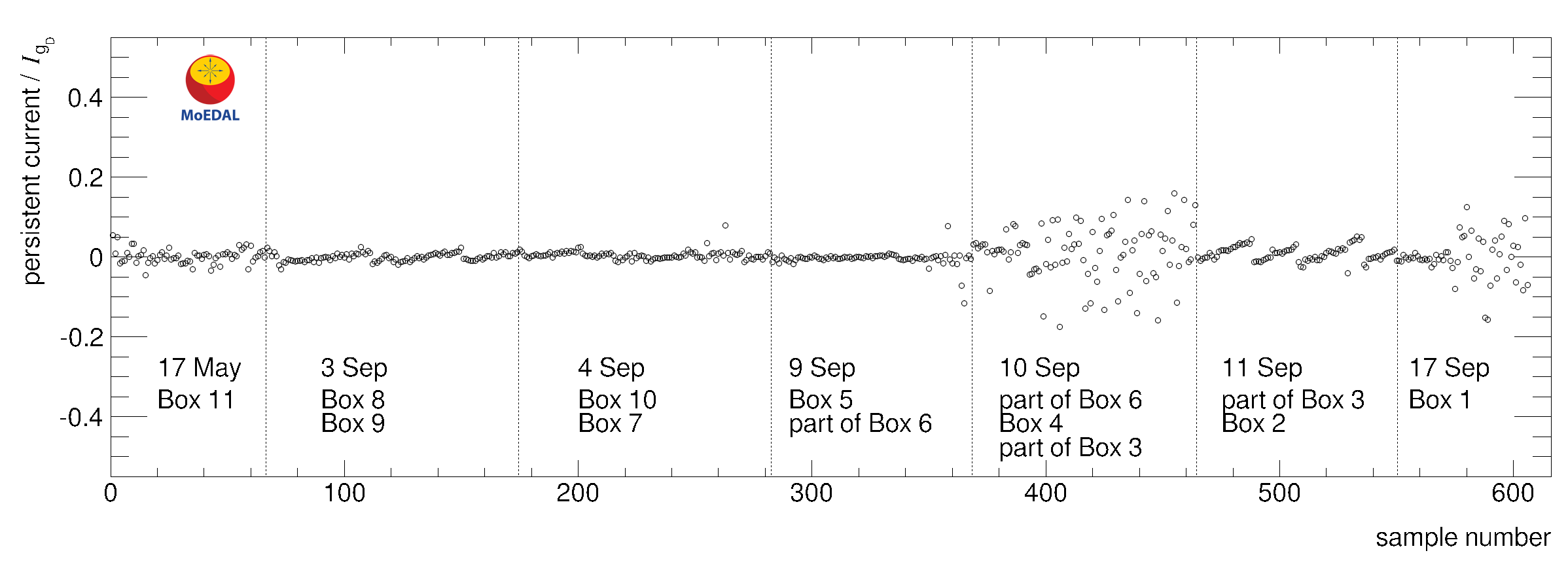}
  \includegraphics[width=0.55\linewidth]{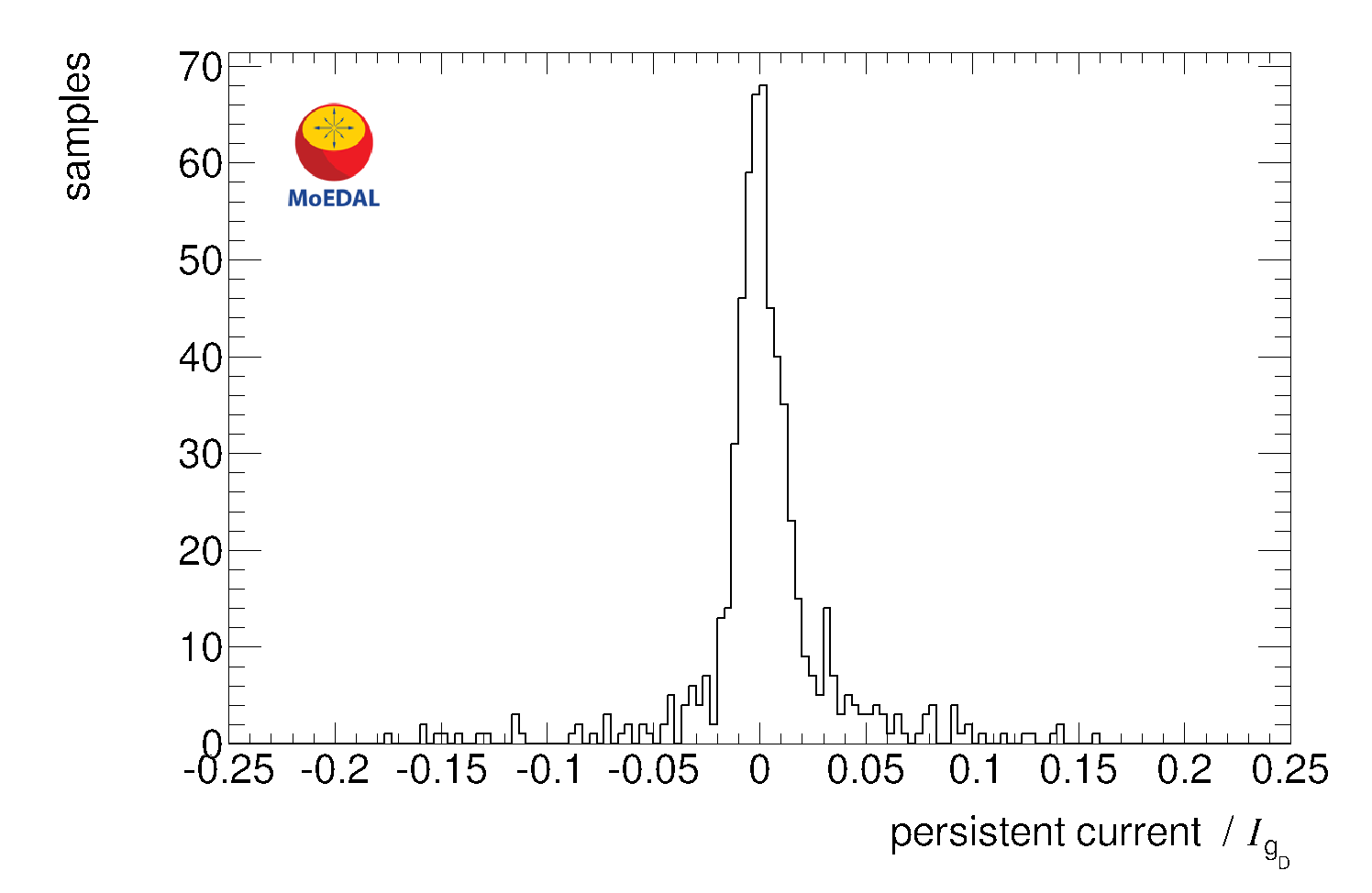}
  \caption{Magnetic charge (in units of the Dirac charge) measured in the 606 aluminium samples of the 2012 MoEDAL trapping detector. The measurements are presented for individual days in the top plot: relative SQUID instabilities can be seen for 10 September and the second half of 17 September, and the saw tooth feature most clearly visible for 11 September is due to offset drifts due to variations of external magnetic fields combined with a scarcity of empty holder measurements. The bottom plot shows the same data as a histogram.}
\label{fig:MMTtest}
\end{center}
\end{figure}

The magnetic charge measurements made for all 606 samples of the trapping detector as measured by the first measurement ---  or the first subsequent measurement in the cases where a spurious offset jump was observed for the first measurement --- are shown in Fig.~\ref{fig:MMTtest}. The upper plot gives an idea of the evolution of the resolution with time, where periods of relative instability are observed for the 10th and 17th of September, which can be attributed to sub-optimal conditions during the measurements for these particular runs. The clearly visible saw tooth feature recorded on the 11th September is due to a relative scarcity of available empty holder measurements, resulting into less frequent offset drift corrections. The bottom plot shows the data as a histogram. No measurements yield values of $|g|$ beyond $0.18g_{\rm D}$. 

The probability that a sample containing a magnetic monopole with $|g|\geq 0.5g_{\rm D}$ would yield a persistent current lower than $0.25g_{\rm D}$, and so remain undetected, was estimated from the rate of spurious jumps with values between $0.25g_{\rm D}$ and $0.5g_{\rm D}$ in a given direction from samples for which multiple measurements confirmed the absence of monopole. This rate was observed to be 0.25\% and is assumed to be the same for samples containing magnetic charge of the order of the Dirac charge because the magnetic charge would result in a current which is small compared to the currents induced by dipoles in the sample and sample holder, which are the main causes for spurious jumps. For monopoles with charges larger than $0.5g_{\rm D}$, this number is even smaller. Thus, the presence of a monopole with $|g|\geq 0.5g_D$ is excluded at more than 99.75\% confidence level. 

\section{Simulation of monopole production and energy loss}
\label{simulations}

\begin{figure}[tb]
\begin{center}
  \includegraphics[width=0.45\linewidth]{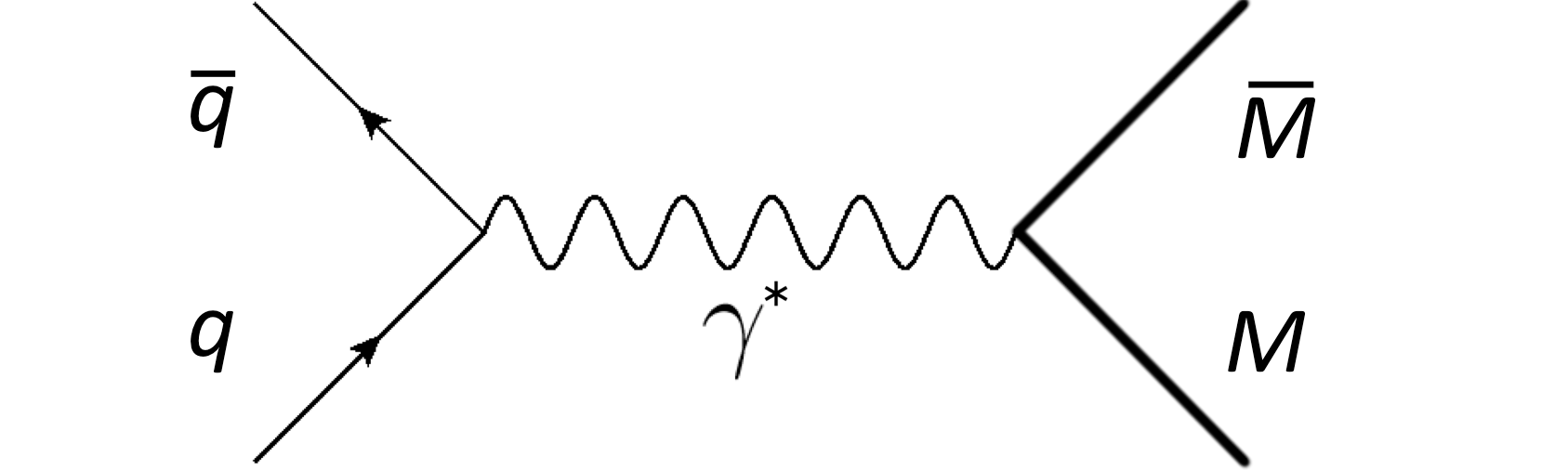}
  \caption{The Feynman diagram for monopole pair production via the Drell-Yan process at leading order. The non-perturbative nature of the process is ignored to allow to provide an interpretation of the search in this simple model.}
\label{fig:Drell-Yan}
\end{center}
\end{figure}

\begin{figure}[tb]
\begin{center}
  \includegraphics[width=0.495\linewidth]{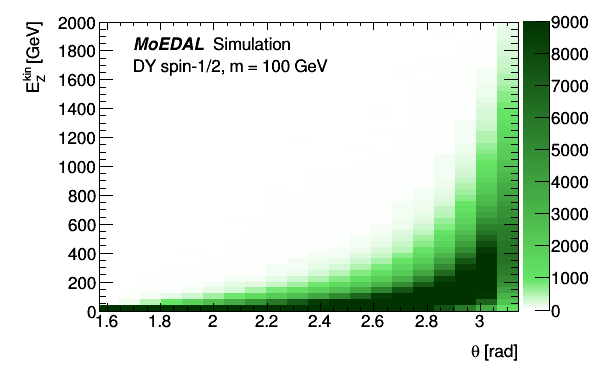}
  \includegraphics[width=0.495\linewidth]{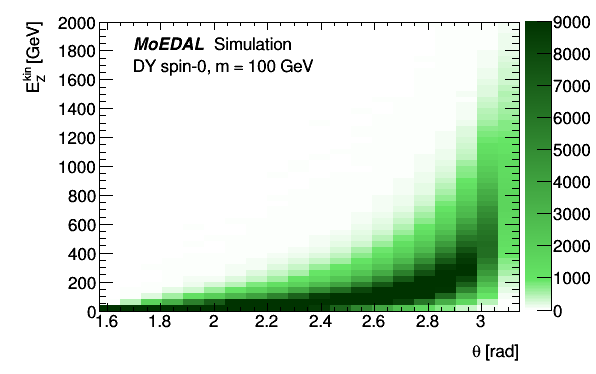}
  \includegraphics[width=0.495\linewidth]{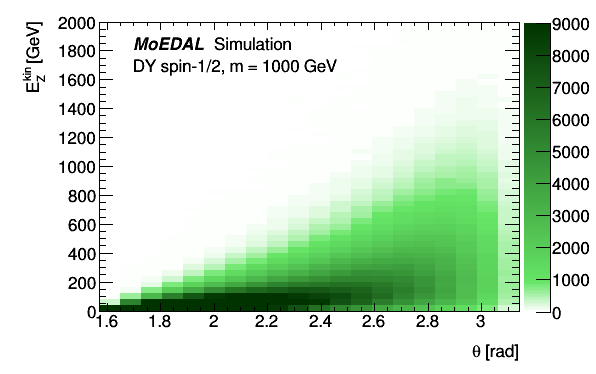}
  \includegraphics[width=0.495\linewidth]{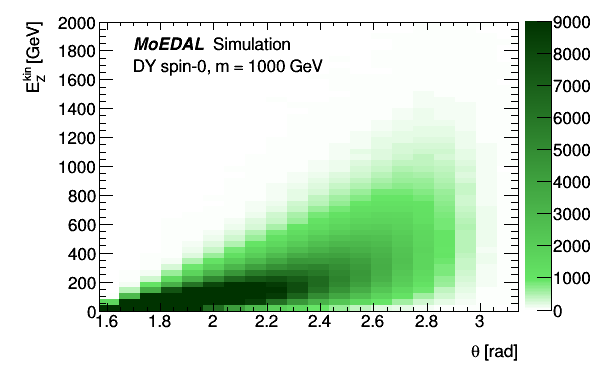}
  \includegraphics[width=0.495\linewidth]{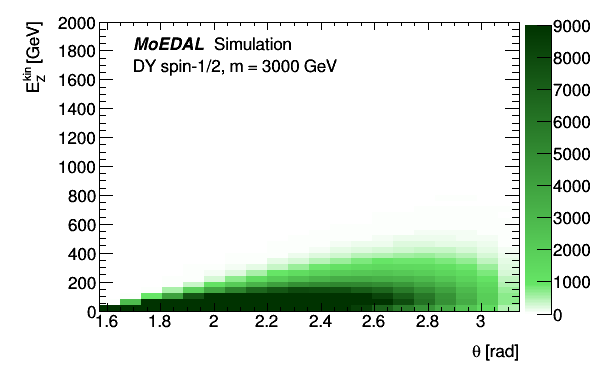}
  \includegraphics[width=0.495\linewidth]{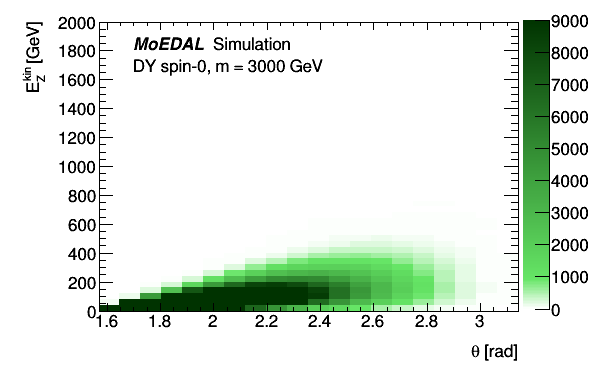}
  \caption{Generator-level distributions of monopoles produced in the Drell-Yan model in the plane described by the longitudinal kinetic energy $E^{kin}_{z}$ and polar angle $\theta$, for spin-1/2 (left) and spin-0 (right) monopoles with mass 100 GeV (top), 1000 GeV (middle) and 3000 GeV (bottom), using $10^{6}$ events for each sample. These distributions do not depend on the monopole charge. The distribution in the range $0 < \theta < \pi/2$ is symmetric to the one that is shown here. }
\label{fig:DYkin}
\end{center}
\end{figure}

Monopole pair production from the initial $pp$ state is modelled by a Drell-Yan (DY) process calculated at leading order, as shown in Fig.~\ref{fig:Drell-Yan}, using the {\sc MadGraph5} Monte-Carlo event generator~\cite{Alwall2014} with the \verb!NNPDF23_lo_as_0130!~\cite{Ball2013} proton parton distribution function, for spin-1/2 and spin-0 monopoles with masses between 100 GeV and 3500 GeV. This model, which ignores the non-perturbative nature of the process due to the large coupling between the monopole and the photon, is chosen by virtue of its simplicity and used to provide an interpretation of the search. Examples of the resulting distributions in the plane described by the longitudinal kinetic energy $E^{kin}_z=E^{kin}\cdot \sin(\theta)$ and polar angle $\theta$ are shown in Fig.~\ref{fig:DYkin}. The use of spin-0 monopoles in addition to spin-1/2, which results in distinctly different DY kinematics due to phase-space constraints resulting from angular momentum conservation, provides an estimate of the model dependence of the detector acceptance. {\sc Pythia~6}~\cite{Sjostrand2006} is used for the initial-state QCD radiation and the hadronisation of the underlying event. Single-monopole samples are also generated with a flat kinetic energy distribution ranging from 0 to 10000 GeV to obtain model-independent results. They are produced with flat $\theta$ and $\phi$ distributions in the ranges corresponding to the angular acceptance of the prototype trapping detector i.e. $2.4$ rad $<\theta<3.0$ rad and $-2.7$ rad $<\phi<-0.5$ rad.

Simulations of monopole propagation and energy loss in the MoEDAL experimental setup are performed using the G\textsc{eant}4 toolkit, with the energy loss behaviour described in Section~\ref{monopole_energy_loss}. The simulations are performed for monopole masses $m$ equal to 100, 500, 1000, 2000, 3000, and 3500 GeV and charges $|g|$ equal to 1, 2, 3, 4, 5 and 6 $g_{\rm D}$, with $2\cdot 10^6$ events in each sample for the single monopoles, and $10^5$ events in each sample for DY production. For the assessment of systematic uncertainties, the simulations are performed using four different geometries, corresponding to the baseline, minimum material, maximum material, and trapping detector position shifted by 1 cm in all three coordinates. For each event, the monopole final position is recorded to assess the acceptance of the trapping detector, or probability to trap a monopole. The results are shown and discussed in Section~\ref{monopole_trapping_acceptance}.

\subsection{Energy loss} 
\label{monopole_energy_loss}

\begin{figure}[tb]
\begin{center}
  \includegraphics[width=0.495\linewidth]{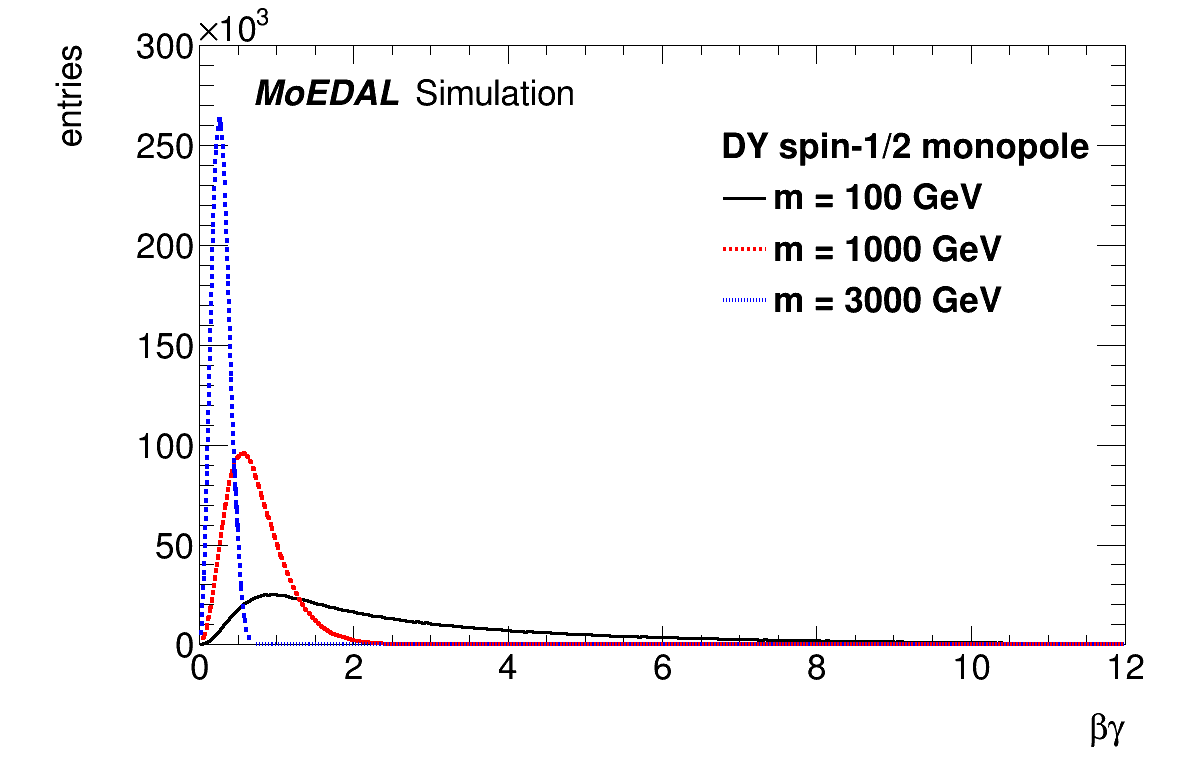}
  \includegraphics[width=0.495\linewidth]{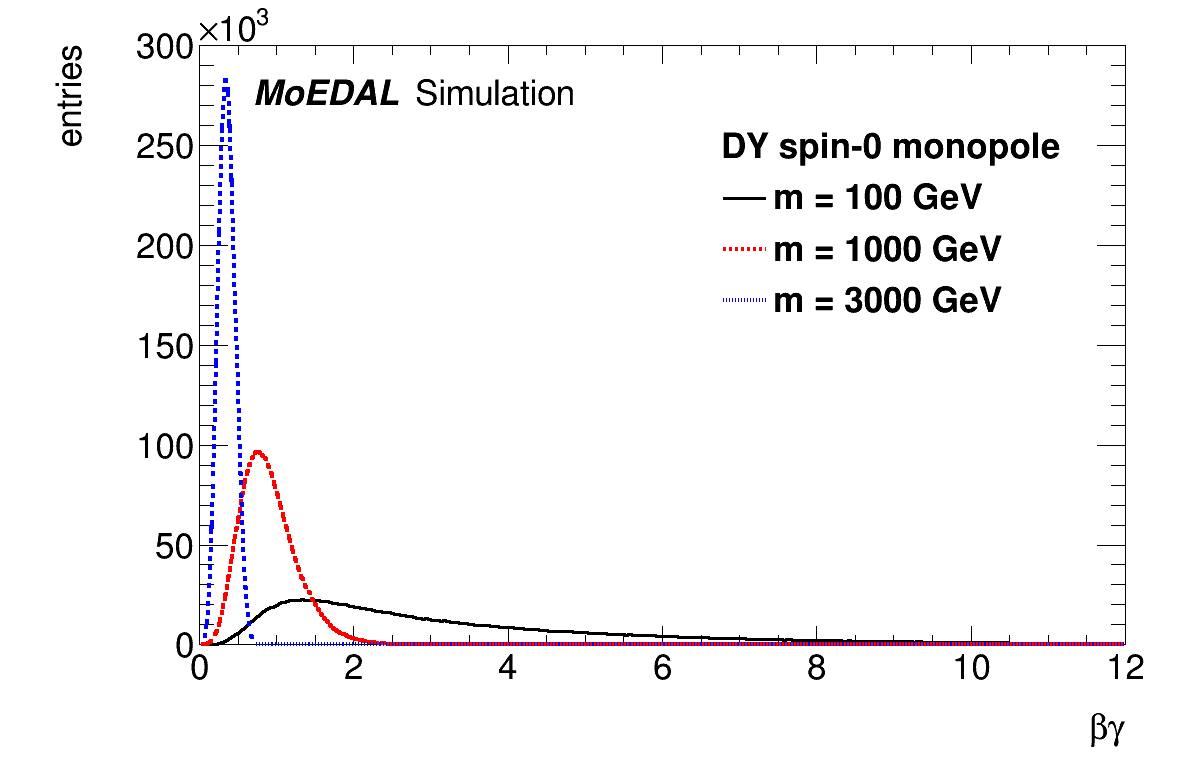}
  \caption{Initial $\beta\gamma$ distributions (with $\beta=v/c$ and $\gamma=\frac{1}{\sqrt{1-\beta^2}}$) of spin-1/2 (left) and spin-0 (right) monopoles produced in the Drell-Yan model with masses in the range considered for this search. These distributions do not depend on the monopole charge.}
\label{fig:DYbetagamma}
\end{center}
\end{figure}

Monopole energy loss in the material inside and around the LHCb VELO vacuum vessel and inside the trapping detector itself (see Section~\ref{experiment}) is simulated using the G\textsc{eant}4 toolkit~\cite{Geant42006}. The velocity dependence of the energy loss per unit distance, for monopoles with velocity $\beta>$0.1 (with $\beta=\frac{v}{c}$), is modelled by the Bethe-Bloch formula modified for monopoles~\cite{Ahlen1978,Derkaoui1998} shown in the following equation, valid for velocities $\beta$ down to $\sim 0.1$:

\begin{equation}
-\frac{\textrm{d}E}{\textrm{d}x}=C\frac{Z}{A}g^2\left[\ln\frac{2m_ec^2\beta^2\gamma^2}{I}+\frac{K(|g|)}{2}-\frac{1}{2}-B(|g|)-\frac{\delta}{2}\right]
\label{Bethe_mag}
\end{equation}
where $Z$, $A$ and $I$ are the atomic number, atomic mass number and mean excitation energy of the medium, $C=\frac{e^4}{m_u4\pi\epsilon_0^2m_ec^2}=0.307$ MeV~g$^{-1}$cm$^2$, $m_u$ is the unified atomic mass unit, $m_e$ is the electron mass and $\gamma=1/\sqrt{1-\beta^2}$. In the monopole velocity range relevant for this search ($\beta\gamma$ generally much lower than 10, as shown in Fig.~\ref{fig:DYbetagamma}), the density correction $\frac{\delta}{2}$ can be neglected in the calculations~\cite{Cecchini2016}.
The Kazama, Yang and Goldhaber cross-section correction and the Bloch correction are given by $K(|g|)=$ 0.406 (0.346) for $|g|=g_D$ ($|g|>g_D$) and $B(|g|)=$ 0.248 (0.672, 1.022, 1.685) for $|g|=g_D$ ($2g_D$, $3g_D$, $6g_D$) \cite{Ahlen1978,Kalbfleisch2004, Derkaoui1998}. The corrections for the intermediate charges $4g_D$ and $5g_D$ are interpolated from these values. Varying $\frac{K(|g|)}{2}-B(|g|)$ by an assumed uncertainty of $\pm 0.5$ results in a 10\% relative uncertainty in d$E$/d$x$, which in turn results in a $1-7$\% uncertainty in the trapping detector acceptance depending on the monopole mass and charge. 

For monopoles with velocity in the range $10^{-4}<\beta<0.01$ the energy losses can be computed assuming that the medium is a degenerate electron gas. Using the result of Ahlen and Kinoshita \cite{Ahlen1982,Craigie1986,Derkaoui1998} an approximation for the energy loss for monopoles is given by:
\begin{equation}
-\frac{1}{\rho}\frac{\textrm{d}E}{\textrm{d}x} = (f_n+f_c) \left(\frac{g}{g_{\rm D}}\right)^{2} \beta ~\frac{GeV}{cm^{2}g}
\label{low_velocity_monopole-dedx}
\end{equation}
\noindent
where $\rho$ is the density of the material through which the monopole is passing and the factors $f_n$ and $f_c$ are dependent on the material and denote the contributions from the non-conduction and conduction electrons, respectively. This formula was implemented in the G\textsc{eant}4 framework for the three dominant materials in the MoEDAL experimental area: aluminium ($f_n=13.7$ and $f_c=80.0$), steel ($f_n=18.9$ and $f_c=28.4$) and copper ($f_n=19.5$ and $f_c=14.5$)~\cite{Cecchini2016}. The energy loss in the intermediate region $0.01<\beta<0.1$ is obtained by linear interpolation. Other interpolation functions were also tried, with negligible impact on the trapping detector acceptance. The relative uncertainty in d$E$/d$x$ obtained with formula~\ref{low_velocity_monopole-dedx} is estimated to be $\pm 30\%$~\cite{Ahlen1982,Craigie1986}, resulting in a $1-9\%$ relative uncertainty in the overall acceptance depending on the monopole mass and charge. 



 
Kinematically, because of the large monopole mass (100 GeV or more) and the limited centre-of-mass energy (8 TeV), the monopoles are not expected to be highly relativistic in MoEDAL. This can be seen in Fig.~\ref{fig:DYbetagamma} for DY produced monopoles. Therefore, monopole  energy losses from bremsstrahlung and pair production, which become significant only for highly relativistic particles ($\beta>0.9999$ or $\beta\gamma>70$)~\cite{PDG2014}, are negligible compared to ionisation. Monopole acceleration along magnetic field lines is implemented in the G\textsc{eant}4 model but it is irrelevant in this case, as the MoEDAL trapping detector is located in a region of the cavern where the fringe magnetic field from LHCb's dipole magnet is negligible (smaller than $5\cdot 10^{-6}$~T) and does not affect monopole trapping or trajectories in any significant way.


\subsection{Trapping criterion}
\label{monopole_trapping_criterion}

Once the monopoles are produced in a collision at the LHC, they will travel through the surrounding material losing energy as described above. They will slow down and likely become trapped in matter by binding to the nucleus of an atom forming the material. 

In the simulations, the final position of a monopole is the point at which its velocity falls to $\beta \le 10^{-3}$. If that position is inside the volume of the prototype trapping detector, the monopole is considered trapped. It was verified that changing the criterion to $\beta \le 10^{-2}$ does not change the result in any significant way. This is expected since $\beta=10^{-2}$ corresponds to a point where a monopole in the mass range considered in this search is expected to further travel at most a few mm in aluminium~\cite{Cecchini2016}. This distance is much smaller than the trapping detector dimensions. 

It is assumed here that the monopole binding to the aluminium trapping volumes is due to the interaction between the magnetic charge of the monopole and the magnetic moment of the nucleus.  Indeed, a few models, summarised in Ref.~\cite{Gamberg2000}, estimate that binding should occur for $^{27}_{13}$Al (100\% natural abundance). The large and positive anomalous magnetic moment of aluminium makes it a good choice for the MoEDAL trapping detector with a predicted monopole-nucleus binding energy in the range $0.5-2.5$ MeV~\cite{Goebel1983,Bracci1984,Olaussen1985,Gamberg2000}. Although the angular-momentum criterion for monopole binding is straightforward it is clear that current theory of monopole binding to nuclear magnetic dipole moments relies on a number of assumptions~\cite{Gamberg2000}. However, the estimated binding energies given above are large and comparable to shell model splittings. Thus, it is reasonable to believe that in the  strong magnetic field in the vicinity of the monopole the nucleus will undergo nuclear rearrangement, in general allowing the monopole to bind to the nuclear, or even the subnuclear, constituents. Also, even a very conservative estimate for the binding energy of 1 eV would give a lifetime of 10~yr~\cite{Gamberg2000,Milton2006}. There is, therefore, good reason to believe that monopoles would be securely captured in MoEDAL's aluminium trapping volumes.



\section{Monopole trapping acceptance}
\label{monopole_trapping_acceptance}

The acceptance of the trapping detector is defined on an event basis as the probability that at least one of the pair-produced monopoles is trapped inside one of the aluminium bars comprising the prototype trapping detector. This acceptance is determined by propagating monopoles with G\textsc{eant}4 through the geometry model defining the disposition of material comprising the prototype trapping detector and the relevant material in its vicinity.

The acceptance is highly dependent on the energy distribution predicted by the model for monopole production and on the material budget between the trapping detector and the production point. Monopoles with low charges and/or high energies tend to punch through the trapping material and are thus better captured in regions where they are slowed down by thicker upstream material. Monopoles with higher charges and/or low energies tend to range out before they reach the trapping detector, and are thus only trapped in regions of low upstream material. Also, for the same charge and the same kinetic energy, monopoles with lower masses possess a higher velocity, leading to higher d$E$/d$x$, and thus tend to lose their energy earlier. For instance, for a magnetic charge $|g|=2g_{\rm D}$, a monopole with mass 100 GeV needs about 100 GeV more kinetic energy to reach the trapping detector than a monopole with mass 1000 GeV.

\begin{figure}[tb]
\begin{center}
  \includegraphics[width=0.495\linewidth]{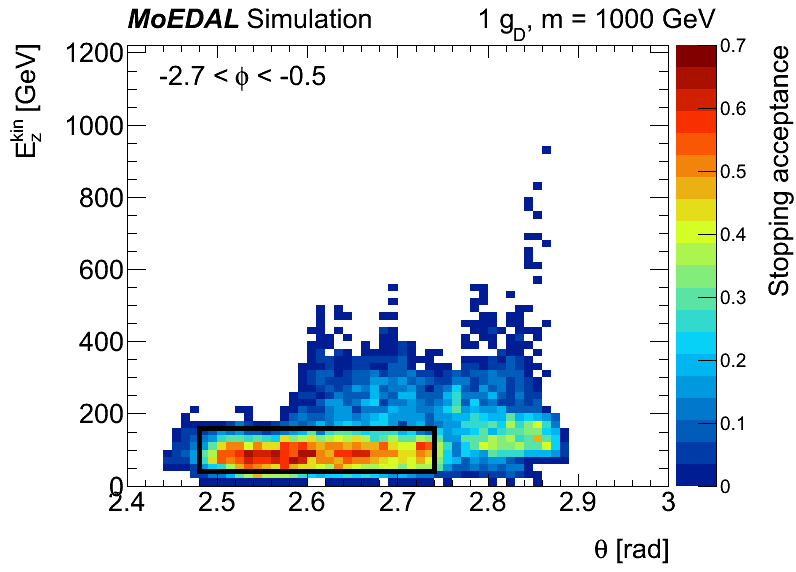}
  \includegraphics[width=0.495\linewidth]{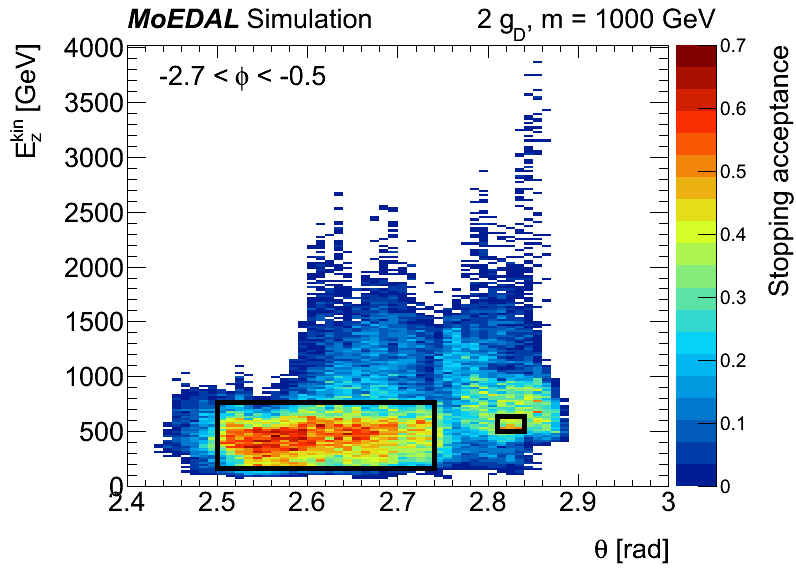}
  \includegraphics[width=0.495\linewidth]{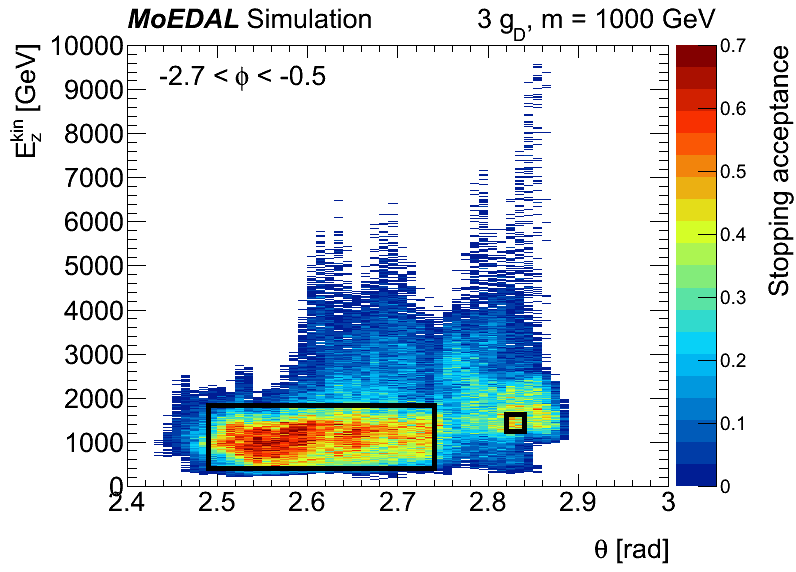}
  \includegraphics[width=0.495\linewidth]{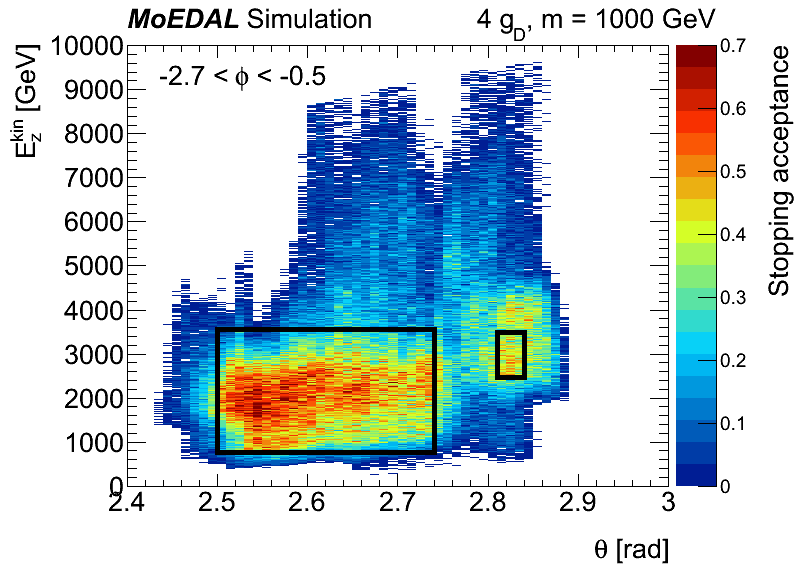}
  \includegraphics[width=0.495\linewidth]{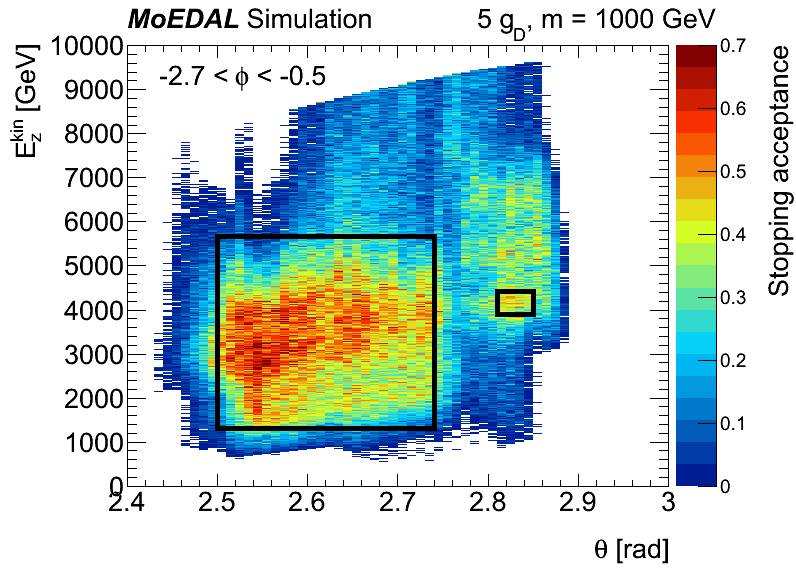}
  \includegraphics[width=0.495\linewidth]{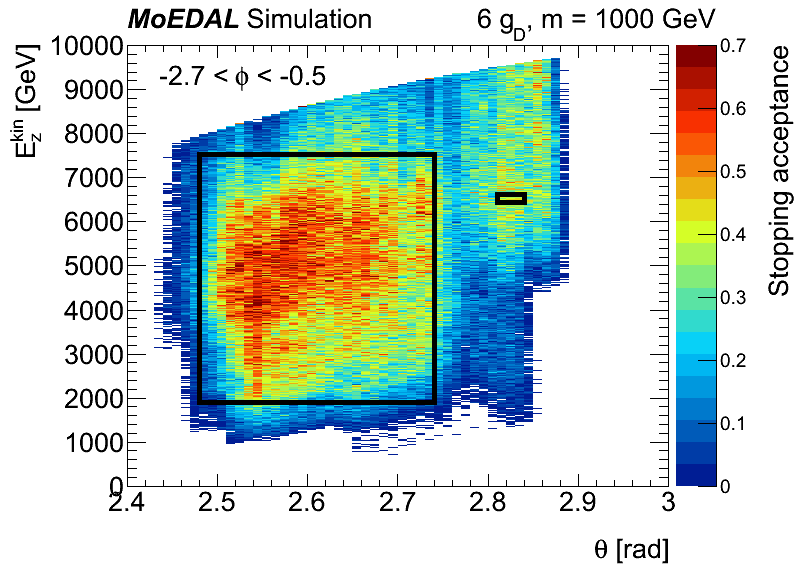}
  \caption{Trapping acceptance as a function of longitudinal kinetic energy $E^{kin}_z$ and polar angle $\theta$  (with $-2.7$ rad $<\phi<-0.5$ rad), for monopoles with mass 1000 GeV and charges ranging from $1g_{\rm D}$ (top, left) to $6g_{\rm D}$ (bottom, right). The fiducial regions (as defined in the text) are indicated by black rectangles.}
\label{fig:accmaps_SP}
\end{center}
\end{figure}

\begin{figure}[tb]
\begin{center}
  \includegraphics[width=0.495\linewidth]{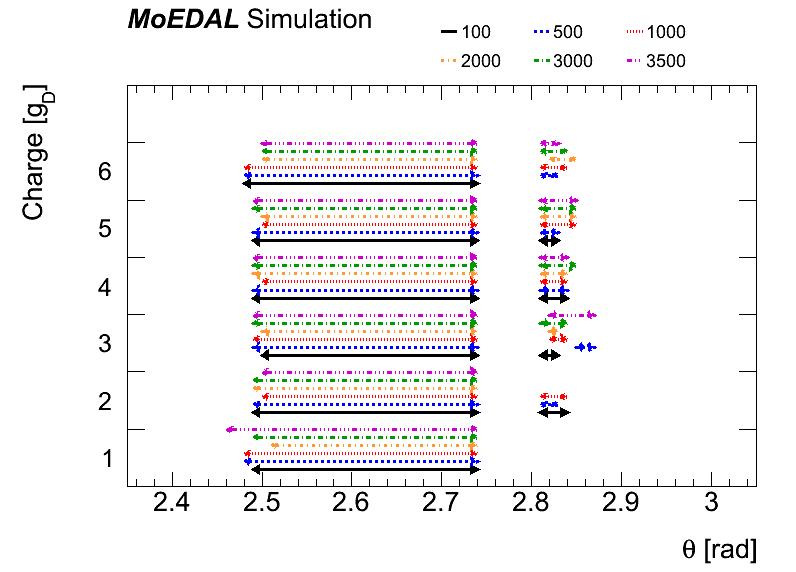}\\
  \includegraphics[width=0.495\linewidth]{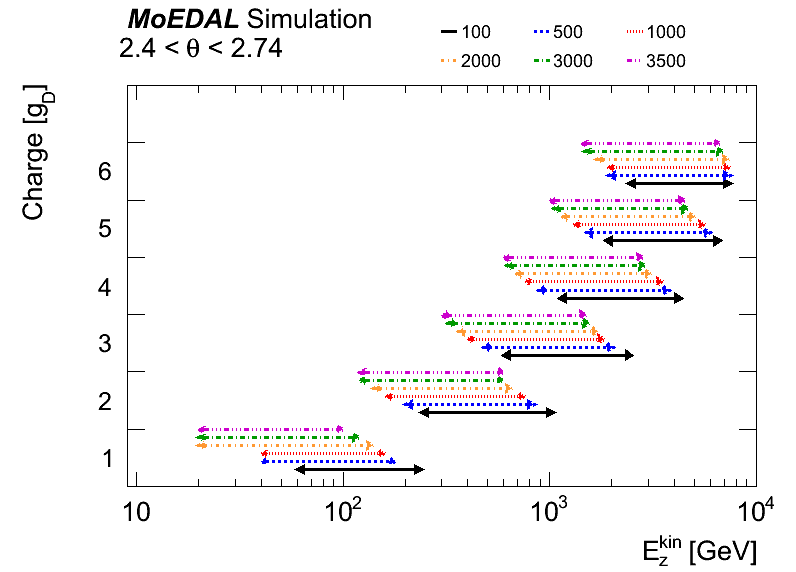}
  \includegraphics[width=0.495\linewidth]{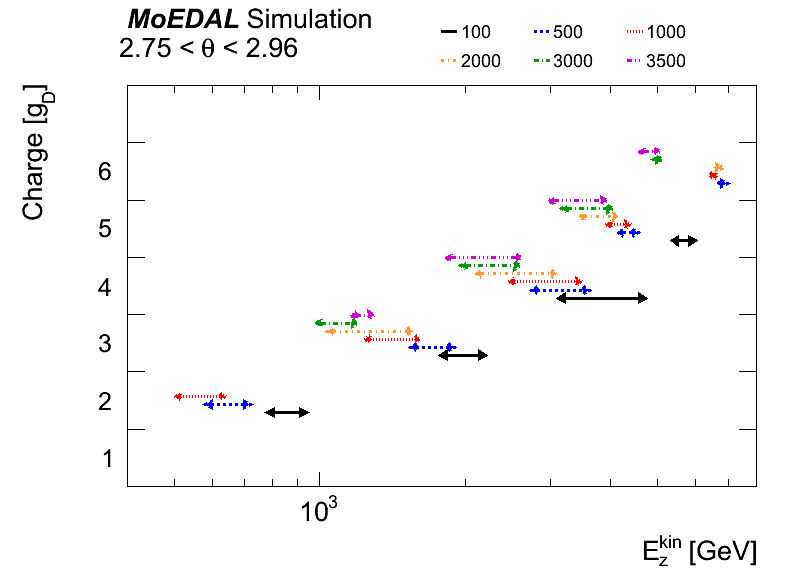}
  \caption{Graphical representation of the fiducial regions for various monopole charges and masses, defined as rectangles in the $\theta$ versus $E^{kin}_z$ plane (with $-2.7$ rad $<\phi<-0.5$ rad) for which the average selection efficiency is larger than 0.4 with a standard deviation lower than 0.15. The double arrows define the rectangle positions and dimensions, with various line styles corresponding to different monopole masses. The top plot shows the $\theta$ acceptance ranges, while the other plots show the $E^{kin}_z$ acceptance ranges corresponding to the two different $\theta$ ranges. The relative uncertainty in the lower and upper $E^{kin}_z$ bounds due to material and d$E$/d$x$ uncertainties is $\pm 25\%$. }
\label{fig:fiducial_regions}
\end{center}
\end{figure}

\subsection{Model-independent analysis}

For a monopole possessing a given charge and mass with a given energy and direction at the origin, the acceptance can be defined in an unique way that depends only on the geometry and not on the production model. Since the collisions are symmetric with respect to the azimuthal angle $\phi$, only two kinematic variables are needed to define the acceptance in a model-independent manner~\cite{ATLAS2011a,ATLAS2012a}. These two variables are chosen here as the longitudinal kinetic energy $E^{kin}_z$ and the polar angle $\theta$. Particle-gun Monte-Carlo samples are used to map the acceptance for all mass and charge combinations as a function of $E^{kin}_z$ and $\theta$ (with $-2.7$ rad $<\phi<-0.5$ rad, corresponding the $\phi$ range of the detector), as shown in Fig.~\ref{fig:accmaps_SP} for monopoles with $m=1000$ GeV. These two-dimensional histograms contain all the information needed to obtain the acceptance for any given pair-production model to a good approximation, for example the Drell-Yan production model (see below). In order to present these data in a clear and simple way (at the cost of some precision, and conservatively neglecting low-acceptance regions) the information can be condensed by considering only the regions in which the acceptance is reasonably high called fiducial regions. An average efficiency of 0.4 is chosen for the fiducial region definition as a trade-off between specifying a large enough region and keeping a uniform acceptance within the region. 

Fiducial regions in the monopole $E^{kin}_z$ versus $\theta$ plane are indicated by black rectangles in Fig.~\ref{fig:accmaps_SP}. These regions are defined, for each charge and mass, by an algorithm that identifies the largest rectangle for which the average selection efficiency between all bins inside the region is larger than 0.4  with a maximum standard deviation of 0.15. Due to the presence of a vacuum pump in front of the upper part of the trapping volume, and also to the fact that the top row comprises only one box instead of two, the acceptance can be divided into two distinct regions in $\theta$, in which rectangles are identified separately: low ($2.40<\theta<2.74$) and high ($2.75<\theta<2.96$) regions. For completeness, the maximum $E^{kin}_z$ is allowed to exceed the beam energy of 4000 GeV minus the monopole mass despite the fact that this can be nonphysical for pair-produced monopoles. 

The ranges of $\theta$ and $E^{kin}_z$ defining the fiducial regions found with this algorithm for all charge and mass points considered in this search (with blank spaces in the cases where no region is found) are summarised graphically in Fig.~\ref{fig:fiducial_regions}, where the top plot shows the intervals in $\theta$, the bottom left plot shows the intervals in $E^{kin}_z$ for the low-$\theta$ region, and the bottom right plot shows the intervals in $E^{kin}_z$ for the high-$\theta$ region.

\subsection{Acceptance for Drell-Yan produced monopoles}

%

DY acceptances are obtained by fully simulating the DY pair-produced monopoles in the geometry model with G\textsc{eant}4. The acceptance depends on the model kinematics as well as the mass and charge of the monopole, as summarised in Table~\ref{tab:eff_models}. It is highest (around 0.02) for charge $2g_{\rm D}$ and intermediate masses. 

The dominant source of systematics is the uncertainty in the estimated amount of material in the geometry description used by the G\textsc{eant}4 simulation. While the VELO vacuum vessel is modelled with great precision in the LHCb geometry, detailed technical drawings of cables and pipes present downstream of the VELO, as well as the interiors of elements such as a  vacuum pump and a vacuum manifold, are not available. Consequently, material comprising these elements was estimated by inspection and direct measurement. Two geometry models are used to obtain estimates of the systematic error due to our imperfect knowledge of the material. These models describe the minimum and maximum amounts of material assuming conservative uncertainties on material thicknesses and densities (see Section~\ref{experiment} for details). This results in a roughly 25\% relative uncertainty in the lower and higher $E^{kin}_z$ boundaries of the fiducial regions used for the model-independent analysis described above (Fig.~\ref{fig:fiducial_regions}). The systematic uncertainty in the material map also results in uncertainties in DY acceptances. With $|g|=g_{\rm D}$, the resulting relative uncertainty is of the order of 10\%. In the case $|g|=2g_{\rm D}$ it is of the order of $10-20\%$ for intermediate masses. The uncertainty is largest for the charge and mass combinations with the lowest acceptance, and exceeds 100\% when the acceptance is below 0.001.

The systematic uncertainty due to the 1 cm uncertainty in the trapping detector position is also estimated. It is found to be in the range $1-17\%$ and is taken into account in the total uncertainty. Another sub-dominant source of systematics is the uncertainty in d$E$/d$x$ in both the low and high $\beta$ regimes, resulting in a $1-10\%$ relative uncertainty in the acceptance as described in Section~\ref{monopole_energy_loss}, also included in the total uncertainty.

The uncertainty resulting from MC statistics lies in the range $1-9\%$ and is always smaller than the total systematic uncertainty.

Trapping detector acceptances for monopoles produced via a DY process, including statistical and systematic uncertainties, are summarised in Table~\ref{tab:eff_models}. This table does not include entries with acceptance lower than 0.001, for which the relative uncertainty exceeds 100\%. This is the case for all DY samples with $|g|>4g_{\rm D}$.

\begin{table}
\begin{center}
\begin{tabular}{|l|c|c|c|c|}
\hline
m [GeV] & $|g|=1.0g_{\rm D}$ & $|g|=2.0g_{\rm D}$ & $|g|=3.0g_{\rm D}$  & $|g|=4.0g_{\rm D}$ \\
\hline
spin-1/2 &&&& \\
100 & 0.019$\pm$0.003 & 0.002$\pm$0.002 & --- & --- \\
500 & 0.017$\pm$0.001 & 0.021$\pm$0.005 & 0.005$\pm$0.003 & --- \\
1000 & 0.014$\pm$0.001 & 0.022$\pm$0.004 & 0.008$\pm$0.004 & 0.002$\pm$0.001 \\
2000 & 0.012$\pm$0.001 & 0.022$\pm$0.003 & 0.008$\pm$0.004 & 0.001$\pm$0.001 \\
3000 & 0.016$\pm$0.001 & 0.013$\pm$0.004 & 0.002$\pm$0.002 & --- \\
3500 & 0.020$\pm$0.001 & 0.004$\pm$0.003 & --- & --- \\
\hline
spin-0 &&&& \\
100 & 0.028$\pm$0.002 & 0.007$\pm$0.004 & --- & --- \\
500 & 0.0082$\pm$0.0010 & 0.027$\pm$0.004 & 0.010$\pm$0.005 & 0.002$\pm$0.002 \\
1000 & 0.0038$\pm$0.0007 & 0.022$\pm$0.002 & 0.011$\pm$0.004 & 0.003$\pm$0.002 \\
2000 & 0.0020$\pm$0.0004 & 0.014$\pm$0.001 & 0.008$\pm$0.003 & 0.002$\pm$0.002 \\
3000 & 0.0032$\pm$0.0007 & 0.008$\pm$0.002 & 0.002$\pm$0.002 & --- \\
3500 & 0.0069$\pm$0.0007 & 0.004$\pm$0.002 & --- & --- \\
\hline
\end{tabular}
\caption{Trapping acceptances for spin-1/2 (top) and spin-0 (bottom) monopoles with DY production kinematic distributions. The quoted absolute uncertainties include both statistical and systematic uncertainties. Empty entries mean that the acceptance is lower than 0.001.}
\label{tab:eff_models}
\normalsize
\end{center}
\end{table}


\section{Limits on monopole production}
\label{limits}

\begin{figure}[tb]
\begin{center}
  \includegraphics[width=0.85\linewidth]{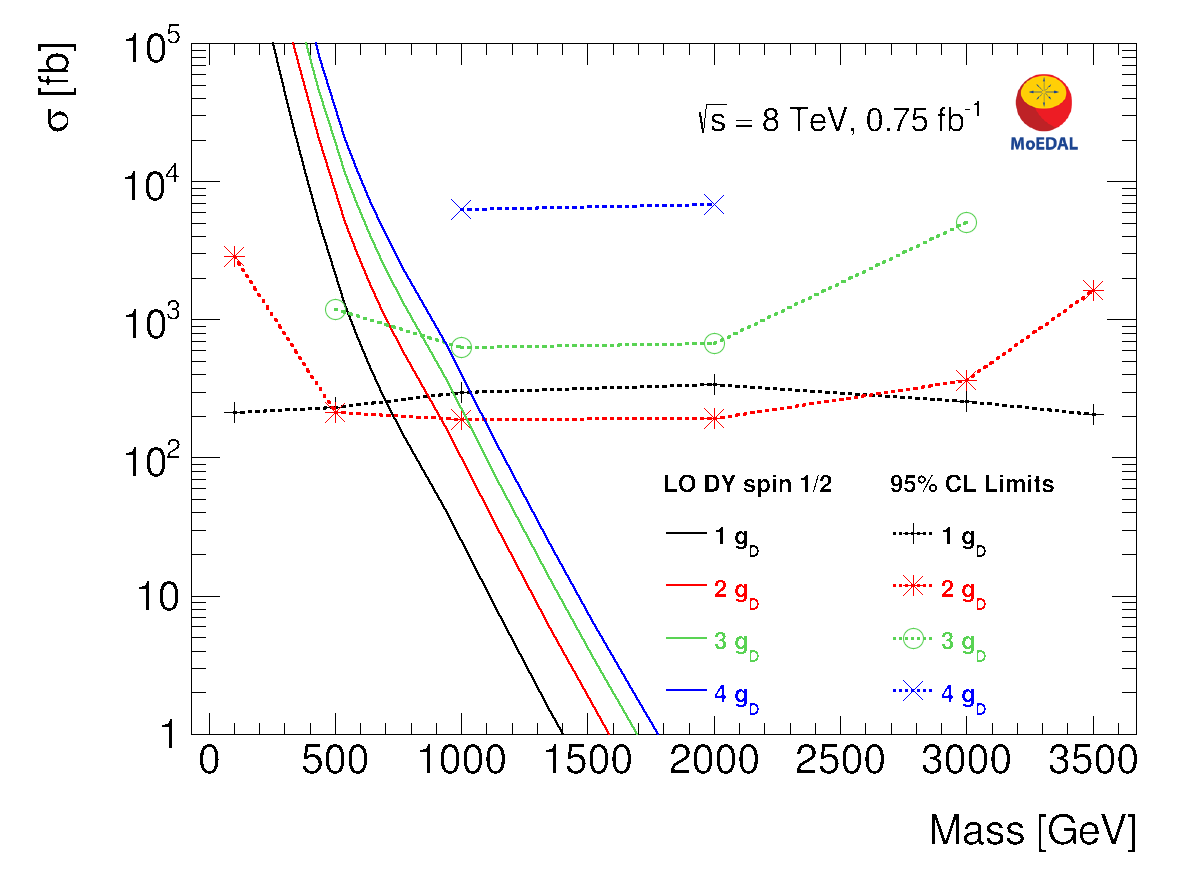}
  \includegraphics[width=0.85\linewidth]{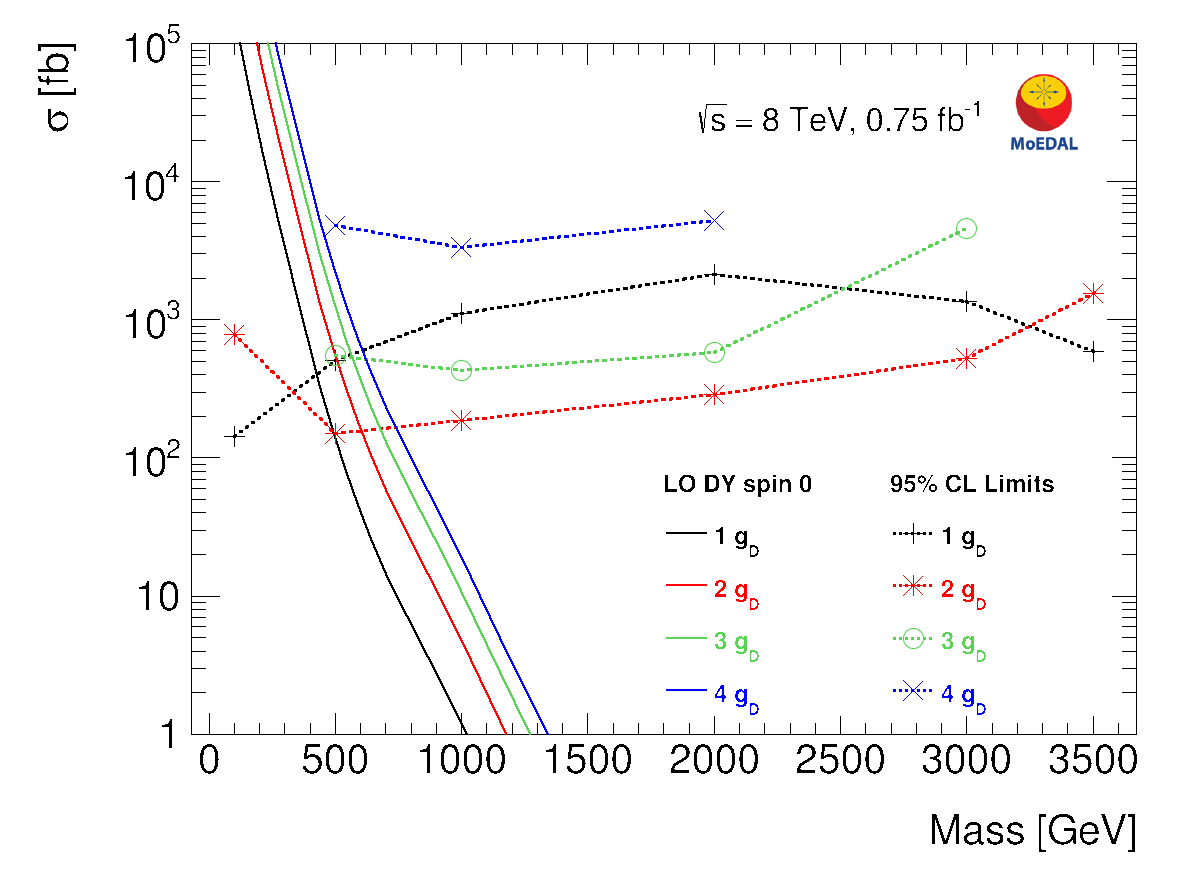}
  \caption{Cross-section upper limits at 95\% confidence level for DY monopole production as a function of mass for spin-1/2 (top) and spin-0 (bottom) monopoles. The various line styles correspond to different monopole charges. The solid lines are DY cross-section calculations at leading order.
}
\label{fig:cross_section_limits_mass}
\end{center}
\end{figure}

\begin{figure}[tb]
\begin{center}
  \includegraphics[width=0.85\linewidth]{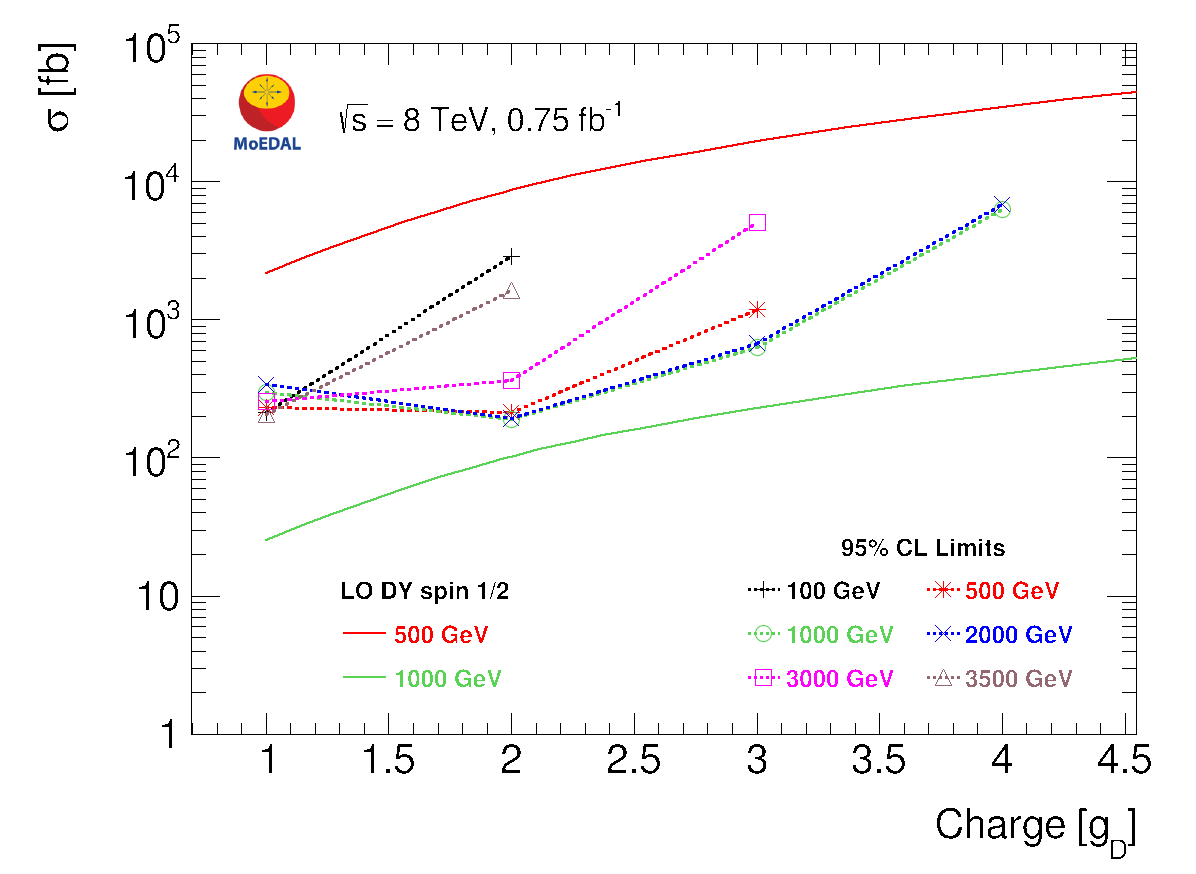}
  \includegraphics[width=0.85\linewidth]{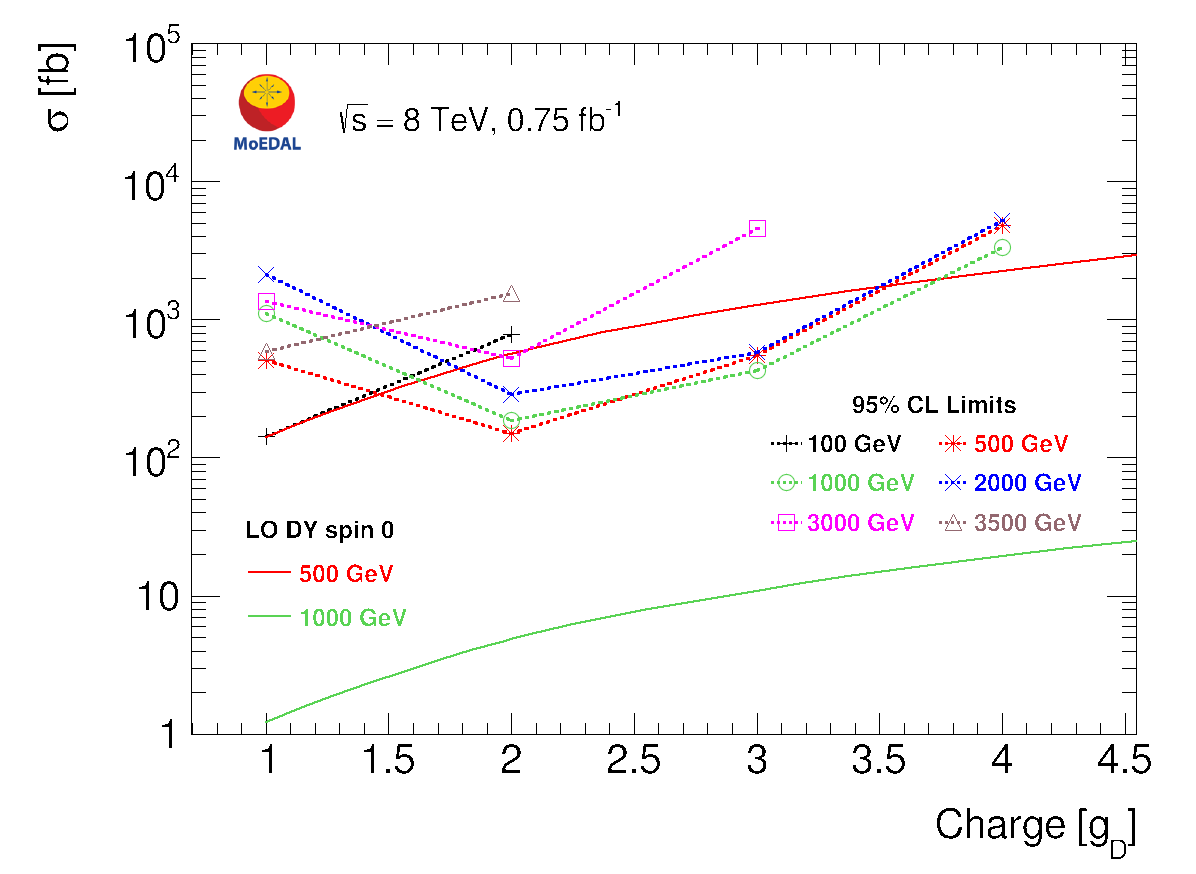}
  \caption{Cross-section upper limits at 95\% confidence level for DY monopole production as a function of charge for spin-1/2 (top) and spin-0 (bottom) monopoles. The various line styles correspond to different monopole masses. The solid lines are DY cross-section calculations at leading order.
}
\label{fig:cross_section_limits_charge}
\end{center}
\end{figure}

No magnetically-charged particles were observed in the prototype trapping detector volumes exposed to $0.75\pm 0.03$ fb$^{-1}$ of 8 TeV $pp$ collisions (Section~\ref{magnetometer}), where the assumption is made that a monopole that would range out in the aluminium material would always be captured and remain bound to a nucleus. This observation, combined with the estimate of the acceptance and its uncertainty (Section~\ref{monopole_trapping_acceptance}), results in stringent limits on the cross sections of processes with monopoles in the final state, using various assumptions for the monopole production kinematics. These limits are obtained using a Bayesian method with Poisson statistics described in detail in Ref.~\cite{Bertram2000}. 

The 95\% confidence level upper limits on the monopole production cross section with the DY process are shown graphically as functions of mass in Fig.~\ref{fig:cross_section_limits_mass} and as a function of charge in Fig.~\ref{fig:cross_section_limits_charge} for spin-1/2 (top) and spin-0 (bottom) monopoles. The DY pair production cross-section calculations are performed at leading order and correspond to the DY cross sections for massive particles with a single electric charge scaled by the factor $g^2 = (n\cdot 68.5)^2$ to account for the monopole charge. These DY cross-section calculations should be viewed with caution since the monopole coupling to the photon is too large for perturbative calculations to converge. However, they are useful as a benchmark by which the present results can be compared with those of other experiments. The corresponding mass limits are shown in Table~\ref{tab:masslimits} for magnetic charges up to $3g_{\rm D}$ for spin-1/2 and spin-0 DY monopoles. The mass limits obtained for $|g|=g_{\rm D}$ are about a factor two lower than the recent ATLAS results at 8 TeV obtained under the same assumptions for monopole production~\cite{ATLAS2015a}. However, mass limits for DY pair produced monopoles with the higher charges $|g|=2g_{\rm D}$ and $|g|=3g_{\rm D}$ are the first to date at the LHC and largely surpass those from previous collider experiments ($\sim 400$ GeV at the Tevatron).

A model-independent result is obtained for monopole production with values of kinetic energy and direction corresponding to the fiducial regions detailed above (see Section~\ref{simulations} and Fig.~\ref{fig:accmaps_SP}). For monopoles within the fiducial regions, a 95\% confidence level upper limit on the cross section of 10 fb is set for monopoles with charges up to 6$g_{\rm D}$ and masses up to 3500 GeV. Again, the superior acceptance of MoEDAL for higher charges is evident compared to previous LHC results~\cite{ATLAS2015a}.


\begin{table}
\begin{center}
\begin{tabular}{|l|c|c|c|}
\hline
  DY Lower Mass Limits [GeV]    & $|g|=g_{\rm D}$ & $|g|=2g_{\rm D}$ & $|g|=3g_{\rm D}$ \\
\hline
spin-1/2  & 700     & 920     & 840   \\
spin-0    & 420     & 600     & 560   \\
\hline
\end{tabular}
\caption{Lower mass limits (95\% confidence level) in models of spin-1/2 (top) and spin-0 (bottom) DY monopole pair production. These limits are based upon cross sections computed at leading order. These cross sections are only indicative since the monopole coupling to the photon is too large to allow for perturbative calculations.}
\label{tab:masslimits}
\end{center}
\end{table}

\section{Conclusions}
\label{conclusions}

The MoEDAL experiment has a pioneering design to search for new physics in the form of highly-ionising particles such as magnetic monopoles or massive (pseudo-)stable charged particles. The largely passive MoEDAL detector, deployed at Interaction Point 8 on the LHC ring, has a dual nature. First, it acts like a giant camera, comprised of nuclear track detectors  sensitive only to new physics. Secondly, it is uniquely able to trap particle messengers of physics beyond the Standard Model for further study. The results of a search for magnetic monopoles utilising a prototype MoEDAL trapping detector exposed to 0.75 fb$^{-1}$ of 8 TeV $pp$ collisions in 2012 --- and subsequently removed and monitored at a remote site by a SQUID magnetometer --- are presented here. This is the first time that a dedicated scalable and reusable trapping array has been deployed at an accelerator facility. 

No monopole candidates with magnetic charge $\geq 0.5g_{\rm D}$ were found in any of the trapping detector samples. Under the assumption of monopole capture by aluminium nuclei, this results in 95\% confidence level cross-section limits ranging from 100 fb to 6000 fb in models of DY monopole pair production for charges up to $4g_{\rm D}$ and masses up to 3500~GeV. Previous LHC constraints for pair production exist only for $|g|\leq 1.5g_{\rm D}$ and $m\leq 2500$~GeV. Under the assumption of a DY cross section at leading order, mass limits are obtained for magnetic charges up to $3g_{\rm D}$. A model-independent 95\% confidence limit of 10 fb is also set for monopoles with charges up to $6g_{\rm D}$ produced in specific ranges of energy and direction.

Despite a small solid angle coverage and modest luminosity, MoEDAL's prototype monopole trapping detector probes ranges of charge, mass and energy which could not be accessed by other LHC experiments. Furthermore, this technique can potentially make a discovery very quickly and allow for an unambiguous background-free assessment of a signal, providing a direct measurement of a monopole magnetic charge based on its magnetic properties only. Importantly,  trapping detectors allow the possibility of studying directly captured magnetic monopoles in the laboratory. A new, larger --- roughly 800 kg --- trapping detector array was deployed in $2015$ downstream of the LHCb VELO vessel as well as on its sides and exposed to $pp$ collisions at 13 TeV centre-of-mass energy. Results from this and other MoEDAL runs will be presented in future publications.

\section*{Acknowledgements}

We thank CERN for the very successful operation of the LHC, as well as the support staff from our institutions without whom MoEDAL could not be operated efficiently. We would like to acknowledge the invaluable assistance of members of the LHCb Collaboration, in particular Guy Wilkinson, Rolf Lindner, Eric Thomas, and Gloria Corti. This work was supported by a fellowship from the Swiss National Science Foundation and a grant from the Marc Birkigt Fund of the Geneva Academic Society; by the London Centre for Terauniverse Studies (LCTS), using funding from the European Research Council via the Advanced Investigator Grant 267352; by the UK Science and Technology Facilities Council (STFC), via the research grants ST/L000326/1, ST/L00044X/1 and ST/N00101X/1; by the Spanish Ministry of Economy and Competitiveness (MINECO), via the grants Grants No. FPA2014-53631-C2-1-P and FPA2015-65652-C4-1-R; by the Generalitat Valenciana via the Projects PROMETEO-II/2013/017 and PROMETEO-II/2014/066, and by the Severo Ochoa Excellence Centre Project SEV-2014-0398; by the Physics Department of King's College London; by a Natural Science and Engineering Research Council of Canada via a project grant; by the V-P Research of the University of Alberta; by the Provost of the University of Alberta; by UEFISCDI (Romania); and by the INFN (Italy).

\bibliographystyle{JHEP}
\bibliography{MMT2015.bib}

\end{document}